# 50+ Metrics for Calendar Mining

Zádor Dániel Kelemen, Dániel Miglász

Kelemen.Daniel@gmail.com, Miglasz.Daniel@gmail.com

**Abstract.** In this report we propose 50+ metrics which can be measured by organizations in order to identify improvements in various areas such as meeting efficiency, capacity planning or leadership skills, just to new a few. The notion of calendar mining is introduced and support is provided for performing the measurement by a reference data model and queries for all metrics defined.

**Keywords:** calendar mining, metric, measure, KPI, event, meeting, calendar, reference data model.

## 1 Introduction

Calendars are usually not analysed, despite that calendar data is available and could be used on personal, team or organizational level for various types of improvements such as supporting decision making, capacity planning, assessing and improving management skills of employees, identifying overloaded employees or to show real work relationships and provide a basis for organizational improvements. Common calendar software such as Microsoft Outlook, Lotus Notes, Google Calendar or Mozilla Lightning do not have features to measure such types of metrics; these tools provide only basic statistics and do not support understanding of calendar data.

We call calendar mining the *processing, analysis and interpretation of calendar data*. Similarly to data mining or process mining [1] and contrast to text mining [2], [3], the input of calendar mining is structured data. Considering this structured data is available, in this report a set of basic metrics is proposed together with their possible usage and an implementation of proposed metrics is presented by introducing a data model and queries for calendar mining.

## 2 Background

Proper ways of organizing meetings has been subject of various research. Content of a meeting, recording, analysing, playback and tagging the content of a meeting is discussed in [4], the internal structure of meetings is addressed in [5]. Based on their practical experiences, start-ups often question the necessity of (recurring) meetings as these are not as much necessary for them as it is preferred by traditional development companies [6]–[8]. Also in the start-up world, there are several initiatives which focus on making meetings better [9] or measure and analyse organizational data[10] in order to gain insight to organizational dynamics.

In contrast to interpretation of unstructured content of meetings or argumentation on the usefulness of meetings, our focus is strictly on options for metrics to be measured on calendar data and providing hints for interpretation of proposed metrics. Section 3 provides the description of our approach, section 4 describes the proposed reference data model, section 5 includes the metrics with hints for interpretation and queries, section 6 provides an overview on limitations, while in section 7 conclusion and further steps are discussed.

## 3 Approach

The focus of this report to answer the following question:

*Can metrics be defined for calendar mining?*

In order to answer the question, 3 main steps were defined:
1. Design a reference data model for storing calendar data
2. Identify calendar metrics
3. Define SQL queries on the reference data model for metrics identified

**Step 1.** Since there are differences among current tools in managing meetings, currently spread calendar data formats do not all support storing all the relevant data, thus it is not straightforward to use these formats as a basis for identifying metrics and writing queries. Therefore, we propose a common and open reference data model which is able to store the relevant data from metrics point of view and can easily be queried and extended.

**Step 2.** As an input for Steps 2 both of the authors spent years in industrial environment, the 1$^{st}$ author serving in various management roles with up to 30 meetings/week. During the analysis of calendars of employees in management role it came out that without measurement and control of meetings, some members of management can become seriously overloaded, having a high number of highly overlapping meetings per week (some employees reached 60 meetings/week in the checked cases). These experiences provided the basic idea of defining meetings related metrics and induced a multi-session brainstorming of authors where metrics were identified. Since brainstorming was used for Step 2, identification of all possible metrics is not assured, nor is intended, rather an initial set provided, which can openly be extended and developed further.

**Step 3.** In order to make the definitions unambiguous, we defined queries for the metrics (identified in step 2) by using the database (introduced in step 1).

# 4 A reference data model for calendar mining

The design of the data model was performed both in an iterative and pragmatic way. Iterations were: 1. Design initial data model for storing most important elements, 2. refine the data model after defining metrics (based on the inputs needed for metrics), 3. refine the data model after defining queries (to make the data model easy to query).

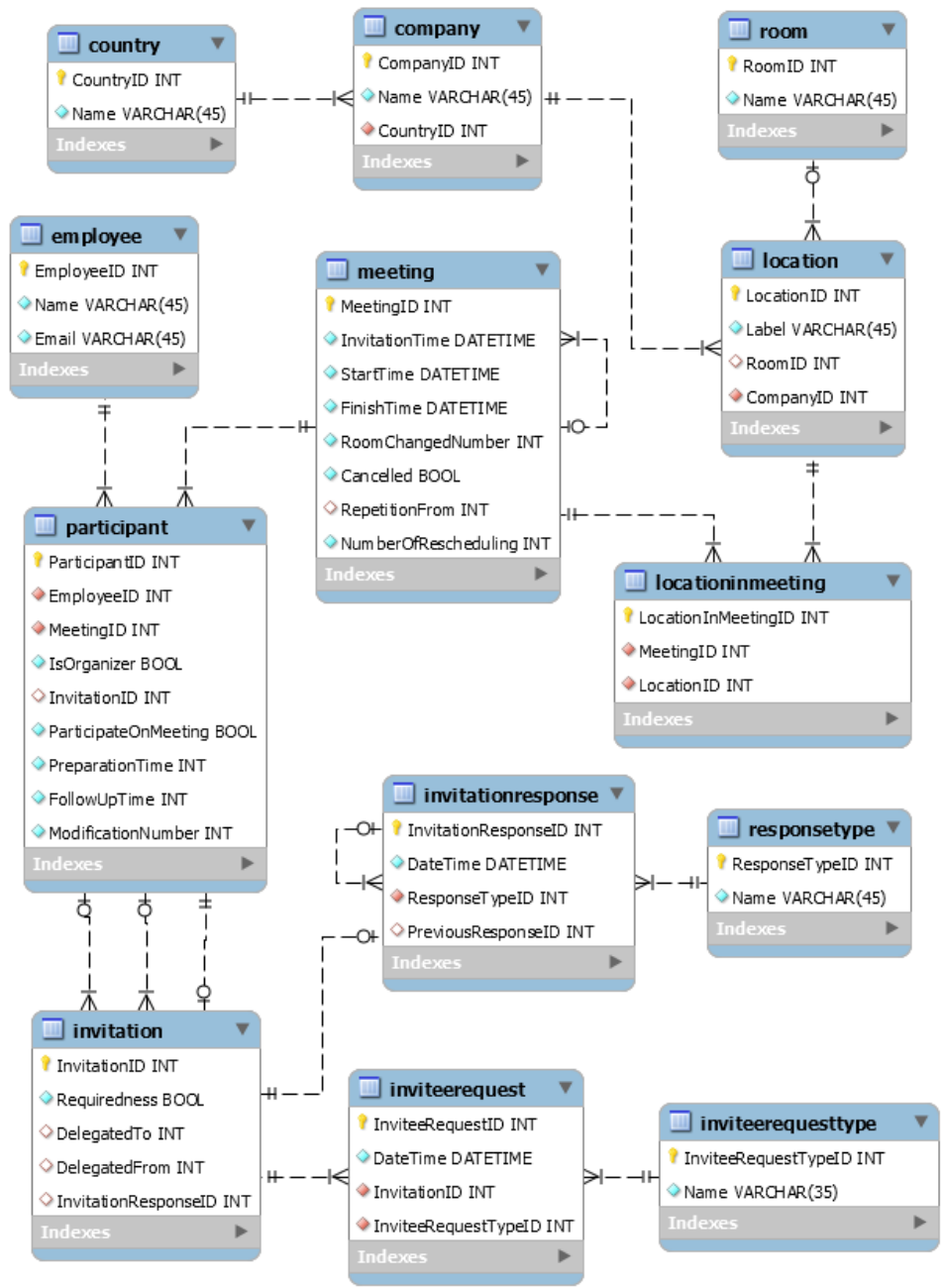

**Figure 1 – A data model for calendar mining**

In order to support easy use and to be pragmatic, the data model was defined not just as an ER model, but by using the free MySQL platform. Figure 1 shows the data model after the third iteration.

The 'meeting' table is the central element of the data model, stores all relevant meeting related data. Location of the meeting is stored in the 'location' table, which points (RoomID) to the 'room' table for storing the rooms. For multi-company meetings, 'company' stores the company at which the meeting is organized, while 'country' stores the country of the meeting. Table 'locationinmeeting' describes the N:M relation of locations and meetings. Table 'employee' stores data of employees, while the 'participant' table connects the tables 'meeting' and 'employee'. Table 'invitation' contains all relevant data related to meeting invitations, while table 'invitationresponse' stores the responses given to meeting invitations. Possible response types to an invitation are stored in the 'responsetype' table.

Table 'inviteerequest' stores requests related to invitations. Possible request types (e.g. information request or re-plan request) are stored in the 'inviteerequestype' table.

Details such as description of attributes of tables can be found in Appendix A – Descriptions of tables in the data model, the SQL script for creating the database in Appendix B - SQL script for creating the database and parameters to be set up before using the queries can be found in Appendix C HELPER: Parameters setting before queries.

## 5   Metrics for calendar mining

In this section metrics to be measured on calendar data are introduced with a name, a definition and a discussion covering their interpretation and possible options of usage. After the discussion of each metric, its related SQL query is presented.

### 5.1   Optional participants

Number of optional participants to be invited to a meeting.
With optional participants the inviter introduces uncertainty both into the meeting (meeting participants will be unsure of the participation of optional participants) and also into the calendars of the optional participants. Even if the optional invitation is accepted optional invitees may think that they can skip the meeting. Therefore measuring number of optional invitations and optional participants can be really useful and optional invitations should be used carefully and reduced if possible.

**List of optional invited participants, who were invited to a specified meeting:**

*@ParamMeetingID = ID of the specified meeting*

```sql
SELECT Name as InvitedParticipants
FROM employee INNER JOIN participant ON employee.EmployeeID = participant.EmployeeID
INNER JOIN invitation ON participant.InvitationID = invitation.InvitationID
WHERE participant.MeetingID = @ParamMeetingID
AND participant.IsOrganizer = 0
AND invitation.Requiredness = 0
```

**Number of optional invited participants, who were invited to a specified meeting**

*@ParamMeetingID = ID of the specified meeting*

```sql
SELECT COUNT(*) as NumberOfInvitedParticipants
FROM participant
INNER JOIN invitation ON participant.InvitationID = invitation.InvitationID
WHERE participant.MeetingID = @ParamMeetingID
AND participant.IsOrganizer = 0
AND invitation.Requiredness = 0
GROUP BY participant.MeetingID
```

**List of numbers of optional invited participants of meetings which were organized by a specified employee (where start time is in a specified interval):**

*@ParamEmployeeID = ID of the specified emplyoee*
*@ParamStartTimeIntervalStart = start time of the specified „start time" interval*
*@ParamStartTimeIntervalEnd = end time of the specified „start time" interval*

```sql
SELECT meetingA.MeetingID, meetingA.StartTime as StartTime,
       IFNULL(InvitedToMeeting.NumberOfInvitedParticipants,0) as NumberOfInvitedParticipants
FROM
meeting meetingA
LEFT JOIN
(
   SELECT COUNT(*) as NumberOfInvitedParticipants, participantB.MeetingID
   FROM participant participantB
   INNER JOIN invitation ON participantB.InvitationID = invitation.InvitationID
   WHERE participantB.IsOrganizer = 0
         AND invitation.Requiredness = 0
   GROUP BY participantB.MeetingID
) InvitedToMeeting
ON meetingA.MeetingID = InvitedToMeeting.MeetingID
WHERE meetingA.MeetingID IN
       (
          SELECT participant.MeetingID
          FROM participant
          WHERE IsOrganizer = 1
                AND participant.EmployeeID = @ParamEmployeeID
       )
       AND meetingA.StartTime >= @ParamStartTimeIntervalStart
       AND meetingA.StartTime <= @ParamStartTimeIntervalEnd
ORDER BY meetingA.StartTime
```

**Average number of optional invited participants of meetings which were organized by a specified employee (where start time is in a specified interval):**

*@ParamEmployeeID = ID of the specified emplyoee*
*@ParamStartTimeIntervalStart = start time of the specified „start time" interval*
*@ParamStartTimeIntervalEnd = end time of the specified „start time" interval*

```sql
SELECT AVG(InvitedToMeeting.NumberOfInvitedParticipants) as AverageNumberOfInvitedParticipants
FROM
meeting meetingA
INNER JOIN
(
   SELECT COUNT(*) as NumberOfInvitedParticipants, participantB.MeetingID
   FROM participant participantB
   INNER JOIN invitation ON participantB.InvitationID = invitation.InvitationID
```

```
   WHERE participantB.IsOrganizer = 0
         AND invitation.Requiredness = 0
   GROUP BY participantB.MeetingID
) InvitedToMeeting
ON meetingA.MeetingID = InvitedToMeeting.MeetingID
WHERE meetingA.MeetingID IN
      (
         SELECT participant.MeetingID
         FROM participant
         WHERE IsOrganizer = 1
               AND participant.EmployeeID = @ParamEmployeeID
      )
      AND meetingA.StartTime >= @ParamStartTimeIntervalStart
      AND meetingA.StartTime <= @ParamStartTimeIntervalEnd
GROUP BY NULL
```

## 5.2   Time between meeting invitation and meeting start

It can happen that invitees receiving meeting invitations sent out too late will not see the invitation or cannot accept it due to other duties. The time needed before invitation depends on multiple parameters (e.g. the meeting type - requires preparation or not, if requires how much time it requires, the invitee physical location - so that they can arrive/join to the meeting). People focusing on their tasks or actively attending other meetings may not check their emails very often, so an appropriate buffer can have positive effects on meeting acceptance.

**Time between invitation and start of a specified meeting**

*@ParamMeetingID = ID of the specified meeting*

```
SELECT TIMEDIFF(meeting.StartTime,meeting.InvitationTime) as TimeBetweenInvitationAndStart
FROM meeting
WHERE meeting.MeetingID = @ParamMeetingID
GROUP BY meeting.MeetingID
```

**List of time between invitation and start of meetings which were organized by a specified employee (where start time is in a specified interval):**

*@ParamEmployeeID = ID of the specified emplyoee*
*@ParamStartTimeIntervalStart = start time of the specified „start time" interval*
*@ParamStartTimeIntervalEnd = end time of the specified „start time" interval*

```
SELECT meeting.MeetingID, meeting.StartTime,TIMEDIFF(meeting.StartTime,meeting.InvitationTime)
as TimeBetweenInvitationAndStart
FROM meeting
WHERE meeting.MeetingID IN
      (
         SELECT participant.MeetingID
         FROM participant
         WHERE IsOrganizer = 1
               AND participant.EmployeeID = @ParamEmployeeID
      )
      AND meeting.StartTime >= @ParamStartTimeIntervalStart
      AND meeting.StartTime <= @ParamStartTimeIntervalEnd
GROUP BY meeting.MeetingID
ORDER BY meeting.StartTime
```

## 5.3 Time between meeting invitation and reaction (accept, reject, tentative, delegate, re-plan request)

Time between meeting invitation and meeting response. Measured on organisational scale this metric can provide a good hint for the minimum time needed between meeting invitation and the meeting itself.

**Last reaction time of a specified meeting**

*@ParamMeetingID = ID of the specified meeting*

```sql
SELECT GREATEST(lastresponse.LastResponse,lastrequest.LastRequest) as lastreaction
FROM
(
    SELECT IFNULL(MAX(invitationresponse.DateTime),0) as LastResponse
    FROM meeting
    INNER JOIN participant ON meeting.MeetingID = participant.MeetingID
    INNER JOIN invitation ON invitation.InvitationID = participant.InvitationID
    LEFT JOIN invitationresponse ON invitationresponse.InvitationResponseID = invitation.InvitationResponseID
    WHERE meeting.MeetingID = @ParamMeetingID
    GROUP BY NULL
    LIMIT 1
) as lastresponse
INNER JOIN
(
    SELECT IFNULL(MAX(inviteerequest.DateTime),0) as LastRequest
    FROM meeting
    INNER JOIN participant ON meeting.MeetingID = participant.MeetingID
    INNER JOIN invitation ON invitation.InvitationID = participant.InvitationID
    LEFT JOIN inviteerequest ON inviteerequest.InvitationID = invitation.InvitationID
    WHERE meeting.MeetingID = @ParamMeetingID
    GROUP BY NULL
    LIMIT 1
) as lastrequest
```

**Time between invitation and last reaction of a specified meeting**

*@ParamMeetingID = ID of the specified meeting*

```sql
SELECT TIMESTAMPDIFF(MINUTE,
    (SELECT InvitationTime FROM meeting WHERE meeting.MeetingID = @ParamMeetingID),
    GREATEST(lastresponse.LastResponse,lastrequest.LastRequest)) as Result
FROM
(
    SELECT IFNULL(MAX(invitationresponse.DateTime),0) as LastResponse
    FROM meeting
    INNER JOIN participant ON meeting.MeetingID = participant.MeetingID
    INNER JOIN invitation ON invitation.InvitationID = participant.InvitationID
    LEFT JOIN invitationresponse ON invitationresponse.InvitationResponseID = invitation.InvitationResponseID
    WHERE meeting.MeetingID = @ParamMeetingID
    GROUP BY NULL
    LIMIT 1
) as lastresponse
INNER JOIN
(
    SELECT IFNULL(MAX(inviteerequest.DateTime),0) as LastRequest
    FROM meeting
    INNER JOIN participant ON meeting.MeetingID = participant.MeetingID
    INNER JOIN invitation ON invitation.InvitationID = participant.InvitationID
    LEFT JOIN inviteerequest ON inviteerequest.InvitationID = invitation.InvitationID
```

```
    WHERE meeting.MeetingID = @ParamMeetingID
    GROUP BY NULL
    LIMIT 1
) as lastrequest
```

## 5.4 Time between meeting response and meeting start

Time between average / last meeting response and meeting start.

**Time between invitation and last response of a specified meeting for every invited employee**

*@ParamMeetingID = ID of the specified meeting*

```
SELECT employee.Name AS Name,
TIMEDIFF(invitationresponse.DateTime,meeting.InvitationTime) AS Time
FROM meeting
INNER JOIN participant ON meeting.MeetingID = participant.MeetingID
INNER JOIN employee ON employee.EmployeeID = participant.EmployeeID
INNER JOIN invitation ON participant.InvitationID = invitation.InvitationID
LEFT JOIN invitationresponse ON invitation.InvitationResponseID =
invitationresponse.InvitationResponseID
WHERE meeting.MeetingID = @ParamMeetingID
AND participant.IsOrganizer = 0
```

**Time between invitation and last response of a specified meeting**

*@ParamMeetingID = ID of the specified meeting*

```
SELECT MIN(TIMEDIFF(invitationresponse.DateTime,meeting.InvitationTime)) AS Time
FROM meeting
INNER JOIN participant ON meeting.MeetingID = participant.MeetingID
INNER JOIN employee ON employee.EmployeeID = participant.EmployeeID
INNER JOIN invitation ON participant.InvitationID = invitation.InvitationID
LEFT JOIN invitationresponse ON invitation.InvitationResponseID =
invitationresponse.InvitationResponseID
WHERE meeting.MeetingID = @ParamMeetingID
AND participant.IsOrganizer = 0
GROUP BY NULL
```

## 5.5 Number of parallel meetings

This metric shows the number of meetings to be attended by a single person per day/week/month. High number of meetings can negatively affect performance of the person or the outcome of meetings the person is invited to, therefore it is important to minimize the number of overlapping meetings.

**Specified two meetings are parallel or not (return 1 if parallel, return 0 if not parallel)**

*@ParamMeetingID1 = ID of the first specified meeting*
*@ParamMeetingID2 = ID of the second specified meeting*

```
SELECT CASE WHEN helper.Result > 0 THEN 1 ELSE 0 END as AreParallel
FROM
(
SELECT
 TIMESTAMPDIFF(MINUTE,
```

```
    GREATEST(
      (SELECT StartTime FROM meeting WHERE MeetingID = @ParamMeetingID1),
      (SELECT StartTime FROM meeting WHERE MeetingID = @ParamMeetingID2)
    ),
    LEAST(
      (SELECT FinishTime FROM meeting WHERE MeetingID = @ParamMeetingID1),
      (SELECT FinishTime FROM meeting WHERE MeetingID = @ParamMeetingID2)
    )) as Result
) as helper
```

**Count of parallel meeting pairs in a specified employee's meetings (where start time is in a specified interval):**

*@ParamEmployeeID = ID of the specified emplyoee*
*@ParamStartTimeIntervalStart = start time of the specified „start time" interval*
*@ParamStartTimeIntervalEnd = end time of the specified „start time" intervalend time of the specified „start time" interval*

```
SELECT COUNT(*) as NumberOfParallelMeetingPairs
FROM
(
SELECT helper.ID1, helper.ID2, helper.TimeBetween
FROM
(
SELECT meetingA.MeetingID as ID1, meetingB.MeetingID as ID2,
  TIMESTAMPDIFF(MINUTE,
    GREATEST(meetingA.StartTime,meetingB.StartTime),
    LEAST(meetingA.FinishTime,meetingB.FinishTime)
  ) as TimeBetween
FROM
(meeting meetingA
INNER JOIN participant participantA ON participantA.MeetingID = meetingA.MeetingID)
CROSS JOIN
(meeting meetingB
INNER JOIN participant participantB ON participantB.MeetingID = meetingB.MeetingID)
WHERE meetingA.MeetingID < meetingB.MeetingID
        AND meetingA.StartTime >= @ParamStartTimeIntervalStart
        AND meetingA.StartTime <= @ParamStartTimeIntervalEnd
        AND meetingB.StartTime >= @ParamStartTimeIntervalStart
        AND meetingB.StartTime <= @ParamStartTimeIntervalEnd
        AND participantA.EmployeeID = @ParamEmployeeID
        AND participantB.EmployeeID = @ParamEmployeeID
        AND participantA.IsOrganizer = 1
        AND participantB.IsOrganizer = 1
) as helper
WHERE helper.TimeBetween > 0
) as helper2
```

## 5.6 Ratio of chaired versus attended meetings

This metric provides the ratio of chaired versus the non-chaired meetings per person. Depending on the organization culture this metrics may indicate leadership role or proactive behaviour of employees.

**Number of chaired and number of attended meeting / person / specified interval**

*@ParamEmployeeID = ID of the specified emplyoee*
*@ParamStartTimeIntervalStart = start time of the specified „start time" interval*
*@ParamStartTimeIntervalEnd = end time of the specified „start time" interval*

```
SELECT participant.IsOrganizer, COUNT(*) as CountParticipant
FROM participant
```

```
INNER JOIN meeting ON meeting.MeetingID = participant.MeetingID
WHERE participant.EmployeeID = @ParamEmployeeID
        AND meeting.StartTime >= @ParamStartTimeIntervalStart
        AND meeting.StartTime <= @ParamStartTimeIntervalEnd
GROUP BY participant.IsOrganizer
```

## 5.7 Response (acceptance/rejection/tentative/no-response) likelihood

Likelihood of the possible responses. It can be counted (1) per meeting or (2) per person. This latter can have several subcases e.g. (2.a) which response (acceptance/rejection/tentative/no-response) is given by what probability by a given person, (2.b) likelihood of answers for a meeting organized by a person, (2.c.1) likelihood of answers of person A reacting to person B's invitations and its opposite: (2.c.2) likelihood of answers of person B reacting to person A's invitations – showing the dynamics between two persons.

**Likelihood of acceptive/tentative/non-responding/decline number of invited participants, who were invited to a specified meeting**

*@ParamMeetingID = ID of the specified meeting*

```
SELECT helper.Response,
        helper.Value / (
                SELECT COUNT(*)
                FROM participant
                INNER JOIN invitation ON participant.InvitationID = invitation.InvitationID
                WHERE participant.MeetingID = @ParamMeetingID
                        AND participant.IsOrganizer = 0
                GROUP BY participant.MeetingID
            ) as Likelihood
FROM (
SELECT 'Accept' as Response,
COALESCE((
   SELECT COUNT(*)
   FROM participant
   INNER JOIN invitation ON participant.InvitationID = invitation.InvitationID
   INNER JOIN invitationresponse ON invitation.InvitationResponseID = invitationresponse.InvitationResponseID
   INNER JOIN responsetype ON invitationresponse.ResponseTypeID = responsetype.ResponseTypeID
   WHERE participant.MeetingID = @ParamMeetingID
   AND participant.IsOrganizer = 0
   AND responsetype.Name = 'accept'
   GROUP BY participant.MeetingID
),0) as Value
UNION ALL
SELECT 'Decline' as Response,
COALESCE((
   SELECT COUNT(*)
   FROM participant
   INNER JOIN invitation ON participant.InvitationID = invitation.InvitationID
   INNER JOIN invitationresponse ON invitation.InvitationResponseID = invitationresponse.InvitationResponseID
   INNER JOIN responsetype ON invitationresponse.ResponseTypeID = responsetype.ResponseTypeID
   WHERE participant.MeetingID = @ParamMeetingID
   AND participant.IsOrganizer = 0
   AND responsetype.Name = 'decline'
   GROUP BY participant.MeetingID
),0) as Value
```

```sql
UNION ALL
SELECT 'Non-Responding' as Response,
COALESCE((
    SELECT COUNT(*)
    FROM participant
    INNER JOIN invitation ON participant.InvitationID = invitation.InvitationID
    WHERE participant.MeetingID = @ParamMeetingID
    AND participant.IsOrganizer = 0
    AND ISNULL(invitation.InvitationResponseID)
    GROUP BY participant.MeetingID
),0) as Value
UNION ALL
SELECT 'Tentative' as Response,
COALESCE((
    SELECT COUNT(*)
    FROM participant
    INNER JOIN invitation ON participant.InvitationID = invitation.InvitationID
    INNER JOIN invitationresponse ON invitation.InvitationResponseID = invitationresponse.InvitationResponseID
    INNER JOIN responsetype ON invitationresponse.ResponseTypeID = responsetype.ResponseTypeID
    WHERE participant.MeetingID = @ParamMeetingID
    AND participant.IsOrganizer = 0
    AND responsetype.Name = 'tentative'
    GROUP BY participant.MeetingID
),0) as Value
UNION ALL
SELECT 'Delegate' as Response,
COALESCE((
    SELECT COUNT(*)
    FROM participant
    INNER JOIN invitation ON participant.InvitationID = invitation.InvitationID
    INNER JOIN invitationresponse ON invitation.InvitationResponseID = invitationresponse.InvitationResponseID
    INNER JOIN responsetype ON invitationresponse.ResponseTypeID = responsetype.ResponseTypeID
    WHERE participant.MeetingID = @ParamMeetingID
    AND participant.IsOrganizer = 0
    AND responsetype.Name = 'delegate'
    GROUP BY participant.MeetingID
),0) as Value
) as helper
```

### 5.8 Meeting rescheduling

Number of times a meeting is rescheduled. Meeting rescheduling could signal several things: the meeting was organized inappropriately multiple times (e.g. because of too many invitees who could not agree or inappropriate timing). Meeting rescheduling could be broken down to further metrics such as number of reschedulings / employee, number of reschedulings / employee versus company, number of rescheduling requests versus number of reschedulings / employee average. These latter three could signal the informal position of the employee within the organisation. No rescheduling could signal leadership or dictating attitude, reschedulings significantly above company average could signal overload of employee or lack of organizing skills.

**Number of rescheduling of a specified meeting:**

*@ParamMeetingID = ID of the specified meeting*

```
SELECT NumberOfRescheduling as NumberOfRescheduling
FROM meeting
WHERE meeting.MeetingID = @ParamMeetingID
```

**List of meetings' rescheduling number which were organized by a specified employee (where start time is in a specified interval):**

*@ParamEmployeeID = ID of the specified emplyoee*
*@ParamStartTimeIntervalStart = start time of the specified „start time" interval*
*@ParamStartTimeIntervalEnd = end time of the specified „start time" interval*

```
SELECT meeting.MeetingID, NumberOfRescheduling as NumberOfRescheduling
FROM meeting
WHERE meeting.MeetingID IN
        (
           SELECT participant.MeetingID
           FROM participant
           WHERE IsOrganizer = 1
                AND participant.EmployeeID = @ParamEmployeeID
        )
        AND meeting.StartTime >= @ParamStartTimeIntervalStart
        AND meeting.StartTime <= @ParamStartTimeIntervalEnd
ORDER BY meeting.MeetingID ASC
```

**Average rescheduling number of meetings which were organized by a specified employee (where start time is in a specified interval):**

*@ParamEmployeeID = ID of the specified emplyoee*
*@ParamStartTimeIntervalStart = start time of the specified „start time" interval*
*@ParamStartTimeIntervalEnd = end time of the specified „start time" interval*

```
SELECT AVG(NumberOfRescheduling) as AverageNumberOfRescheduling
FROM meeting
WHERE meeting.MeetingID IN
        (
           SELECT participant.MeetingID
           FROM participant
           WHERE IsOrganizer = 1
                AND participant.EmployeeID = @ParamEmployeeID
        )
        AND meeting.StartTime >= @ParamStartTimeIntervalStart
        AND meeting.StartTime <= @ParamStartTimeIntervalEnd
GROUP BY NULL
```

### 5.9  Invitee modifications

Number of new meeting invitees/invitee removals/changing invitation level of invitees (required-optional-FYI) per meeting or per chair. Modifying invitees may signal that the chair is unsure about the needed participants, this could directly affect the outcome of the meeting. In addition, invitee additions and removals may unnecessarily disturb employees (removed ones will be bothered unnecessarily, added ones will have shorter time between the receiving the invitation and meeting start compared to those invited initially).

**List of invitees' modifications numbers in a specified meeting**

*@ParamMeetingID = ID of the specified meeting*

```sql
SELECT employee.Name as EmployeeName, participant.ModificationNumber as ModificationNumber
FROM participant INNER JOIN employee ON employee.EmployeeID = participant.EmployeeID
WHERE participant.MeetingID = @ParamMeetingID AND participant. IsOrganizer = 0
```

**Sum of invitees' modifications number in a specified meeting**

*@ParamMeetingID = ID of the specified meeting*

```sql
SELECT SUM(participant.ModificationNumber) as SumOfModificationNumber
FROM participant
WHERE participant.MeetingID = @ParamMeetingID AND participant.IsOrganizer = 0
GROUP BY participant.MeetingID
```

**List of invitees' modifications numbers in meetings which were organized by a specified employee (where start time is in a specified interval):**

*@ParamEmployeeID = ID of the specified emplyoee*
*@ParamStartTimeIntervalStart = start time of the specified „start time" interval*
*@ParamStartTimeIntervalEnd = end time of the specified „start time" intervalend time of the specified „start time" interval*

```sql
SELECT meeting.MeetingID, SUM(participant.ModificationNumber) as sumOfModificationNumbers
FROM participant INNER JOIN meeting ON meeting.MeetingID = participant.MeetingID
WHERE meeting.MeetingID IN
        (
            SELECT participant.MeetingID
            FROM participant
            WHERE IsOrganizer = 1
                    AND participant.EmployeeID = @ParamEmployeeID
        )
        AND meeting.StartTime >= @ParamStartTimeIntervalStart
        AND meeting.StartTime <= @ParamStartTimeIntervalEnd
GROUP BY participant.MeetingID
ORDER BY meeting.MeetingID
```

**Average modifications numbers in meetings which were organized by a specified employee (where start time is in a specified interval):**

*@ParamEmployeeID = ID of the specified emplyoee*
*@ParamStartTimeIntervalStart = start time of the specified „start time" interval*
*@ParamStartTimeIntervalEnd = end time of the specified „start time" intervalend time of the specified „start time" interval*

```sql
SELECT AVG(helper. SumOfModificationNumber) as AverageModificationNumber
FROM
(
    SELECT SUM(participant.ModificationNumber) as SumOfModificationNumber
    FROM participant INNER JOIN meeting ON meeting.MeetingID = participant.MeetingID
    WHERE meeting.MeetingID IN
            (
                SELECT participant.MeetingID
                FROM participant
                WHERE IsOrganizer = 1
                        AND participant.EmployeeID = @ParamEmployeeID
            )
            AND meeting.StartTime >= @ParamStartTimeIntervalStart
            AND meeting.StartTime <= @ParamStartTimeIntervalEnd
    GROUP BY participant.MeetingID
) as helper
GROUP BY NULL
```

## 5.10 Meeting length

Minutes planned to be spent on a meeting. Too short or too long meetings without breaks could reduce efficiency. For example it could be inefficient to plan a meeting for 5 minutes or for 5 hours in several cases. A minimum meeting length could be predicted based on parameters such as locations to be involved or number of participants. Similarly too long meetings e.g. 3-4-5 hours without breaks could be exhausting. An intelligent tool could suggest breaks e.g. for a 3 hours meeting 10 minutes per hour.

A hint for meeting length could be incorporated into calendars based on number of locations, number of participants or number of topics to be discussed.

**Length of a specified meeting (in minutes):**

*@ParamMeetingID = ID of the specified meeting*

```
SELECT TIMESTAMPDIFF(MINUTE,meeting.StartTime,meeting.FinishTime) as Length
FROM meeting
WHERE meeting.MeetingID = @ParamMeetingID
```

**List of meetings' length which were organized by a specified employee (where start time is in a specified interval):**

*@ParamEmployeeID = ID of the specified emplyoee*
*@ParamStartTimeIntervalStart = start time of the specified „start time" interval*
*@ParamStartTimeIntervalEnd = end time of the specified „start time" interval*

```
SELECT meeting.MeetingID, meeting.StartTime,
TIMESTAMPDIFF(MINUTE,meeting.StartTime,meeting.FinishTime) as Length
FROM meeting
WHERE meeting.MeetingID IN
        (
            SELECT participant.MeetingID
            FROM participant
            WHERE IsOrganizer = 1
                AND participant.EmployeeID = @ParamEmployeeID
        )
        AND meeting.StartTime >= @ParamStartTimeIntervalStart
        AND meeting.StartTime <= @ParamStartTimeIntervalEnd
ORDER BY meeting.StartTime ASC
```

**Average length of meetings which were organized by a specified employee (where start time is in a specified interval):**

*@ ParamEmployeeID = ID of the specified emplyoee*

*@ParamStartTimeIntervalStart = start time of the specified „start time" interval*

*@ParamStartTimeIntervalEnd = end time of the specified „start time" interval*

```
SELECT AVG(TIMESTAMPDIFF(MINUTE,meeting.StartTime,meeting.FinishTime)) as Length
FROM meeting
WHERE meeting.MeetingID IN
        (
            SELECT participant.MeetingID
            FROM participant
            WHERE IsOrganizer = 1
                AND participant.EmployeeID = @ParamEmployeeID
        )
        AND meeting.StartTime >= @ParamStartTimeIntervalStart
        AND meeting.StartTime <= @ParamStartTimeIntervalEnd
GROUP BY NULL
```

## 5.11 Meeting reoccurrences

Number of reoccurrences of a meeting. Meeting reoccurrences can depend on purpose (e.g. reoccurrences of a project meetings could end when the project finishes). Reoccurrences of organisational meetings (such as reviews by the higher level management) could be theoretically infinite since these are to be organized in the whole lifecycle of an organisation, however since most organisations are continuously changing, a maximum of reoccurences for these meetings could also be set (e.g. no longer than 1 year). Periodic reorganisation of meetings with multiple occurences would help rethinking if these meetings are still needed with the same topics, participants, length and frequency.

**Reoccurrences of a specified meeting**

*@ParamMeetingID = ID of the specified meeting*

```
SELECT COUNT(*) as Reoccurrences
FROM meeting
WHERE
  meeting.RepetitionFrom = (SELECT meeting.RepetitionFrom
                            FROM meeting
                            WHERE meeting.MeetingID = @ParamMeetingID
                            LIMIT 1)
GROUP BY NULL
```

## 5.12 Meeting reoccurrence frequency

Time between two instances of a repetitive meeting. Meetings can be reoccurring with predefined frequency (e.g. daily stand-ups / weekly project meetings/ bi-weekly sprints, monthly reviews by higher level management etc.).
It could be useful to plan rarely happening meetings (e.g. monthly reoccurring meetings a year) in advance, so that invitees can plan their tasks and holidays accordingly. Revisiting reoccurring meetings after a period could also be a useful practice (e.g. deciding on continuation/transformation of a weekly meeting after half a year). Frequently reoccurring meetings may be planned/re-planned in a more dynamic way.

**Reoccurrences frequency of a specified meeting**

**Time difference between first and last occurences / (number of occurences – 1)**

If the meeting has one occurence (there are no repetitions) the result value is NULL.

*@ParamMeetingID = ID of the specified meeting*

```
SELECT TIMESTAMPDIFF(MINUTE,temp.mindate,temp.maxdate)/(temp.Reoccurrences-1) as
Frequency
FROM
(
   SELECT COUNT(*) as Reoccurrences, MIN(meeting.StartTime) as mindate,
          MAX(meeting.StartTime) as maxdate
   FROM meeting
```

```
    WHERE
      meeting.RepetitionFrom = (SELECT meeting.RepetitionFrom
                                FROM meeting
                                WHERE meeting.MeetingID = @ParamMeetingID
                                LIMIT 1)
    GROUP BY NULL
) as temp
```

## 5.13 Invited meeting participants

Number of all participants to be invited to a meeting. This can include groups and various levels of invitations such as (1) required, (2) optional and (3) invitations delivered only for information purposes.

**List of invited participants, who were invited to a specified meeting**

*@ParamMeetingID = ID of the specified meeting*

```
SELECT Name as InvitedParticipants
FROM employee INNER JOIN participant ON employee.EmployeeID = participant.EmployeeID
WHERE participant.MeetingID = @ParamMeetingID
AND participant.IsOrganizer = 0
AND NOT ISNULL(participant.InvitationID)
```

**Number of invited participants, who were invited to a specified meeting**

*@ParamMeetingID = ID of the specified meeting*

```
SELECT COUNT(*) as NumberOfInvitedParticipants
FROM participant
WHERE participant.MeetingID = @ParamMeetingID
AND participant.IsOrganizer = 0
AND NOT ISNULL(participant.InvitationID)
GROUP BY participant.MeetingID
```

**List of numbers of invited participants of meetings which were organized by a specified employee (where start time is in a specified interval)**

*@ParamEmployeeID = ID of the specified emplyoee*
*@ParamStartTimeIntervalStart = start time of the specified „start time" interval*
*@ParamStartTimeIntervalEnd = end time of the specified „start time" interval*

```
SELECT meetingA.MeetingID, meetingA.StartTime as StartTime,
        IFNULL(InvitedToMeeting.NumberOfInvitedParticipants,0) as NumberOfInvitedParticipants
FROM
meeting meetingA
LEFT JOIN
(
   SELECT COUNT(*) as NumberOfInvitedParticipants, participantB.MeetingID
   FROM participant participantB
   WHERE participantB.IsOrganizer = 0
         AND NOT ISNULL(participantB.InvitationID)
   GROUP BY participantB.MeetingID
) InvitedToMeeting
ON meetingA.MeetingID = InvitedToMeeting.MeetingID
WHERE meetingA.MeetingID IN
        (
          SELECT participant.MeetingID
          FROM participant
          WHERE IsOrganizer = 1
                AND participant.EmployeeID = @ParamEmployeeID
        )
```

```
        AND meetingA.StartTime >= @ParamStartTimeIntervalStart
        AND meetingA.StartTime <= @ParamStartTimeIntervalEnd
ORDER BY meetingA.StartTime
```

**Average number of invited participants of meetings which were organized by a specified employee (where start time is in a specified interval)**

*@ParamEmployeeID = ID of the specified emplyoee*
*@ParamStartTimeIntervalStart = start time of the specified „start time" interval*
*@ParamStartTimeIntervalEnd = end time of the specified „start time" interval*

```
SELECT AVG(InvitedToMeeting.NumberOfInvitedParticipants) as AverageNumberOfInvitedParticipants
FROM
meeting meetingA
INNER JOIN
(
    SELECT COUNT(*) as NumberOfInvitedParticipants, participantB.MeetingID
    FROM participant participantB
    WHERE participantB.IsOrganizer = 0
            AND NOT ISNULL(participantB.InvitationID)
    GROUP BY participantB.MeetingID
) InvitedToMeeting
ON meetingA.MeetingID = InvitedToMeeting.MeetingID
WHERE meetingA.MeetingID IN
        (
            SELECT participant.MeetingID
            FROM participant
            WHERE IsOrganizer = 1
                    AND participant.EmployeeID = @ParamEmployeeID
        )
        AND meetingA.StartTime >= @ParamStartTimeIntervalStart
        AND meetingA.StartTime <= @ParamStartTimeIntervalEnd
GROUP BY NULL
```

## 5.14 Required participants

Number of required participants to be invited to a meeting. Number of required participants can highly vary depending on the meeting type. E.g. on meetings such as company information days many people (e.g. the whole company) could be invited. For a more productive or co-working meeting (e.g. within a project) number of required participants may be useful to be limited to keep the meeting productive.

**List of required invited participants, who were invited to a specified meeting**

*@ParamMeetingID = ID of the specified meeting*

```
SELECT Name as InvitedParticipants
FROM employee INNER JOIN participant ON employee.EmployeeID = participant.EmployeeID
INNER JOIN invitation ON participant.InvitationID = invitation.InvitationID
WHERE participant.MeetingID = @ParamMeetingID
AND participant.IsOrganizer = 0
AND invitation.Requiredness = 1
```

**Number of required invited participants, who were invited to a specified meeting**

*@ParamMeetingID = ID of the specified meeting*

```
SELECT COUNT(*) as NumberOfInvitedParticipants
FROM participant
INNER JOIN invitation ON participant.InvitationID = invitation.InvitationID
WHERE participant.MeetingID = @ParamMeetingID
```

```
AND participant.IsOrganizer = 0
AND invitation.Requiredness = 1
GROUP BY participant.MeetingID
```

**List of numbers of required invited participants of meetings which were organized by a specified employee (where start time is in a specified interval)**

*@ParamEmployeeID = ID of the specified emplyoee*
*@ParamStartTimeIntervalStart = start time of the specified „start time" interval*
*@ParamStartTimeIntervalEnd = end time of the specified „start time" interval*

```
SELECT meetingA.MeetingID, meetingA.StartTime as StartTime,
       IFNULL(InvitedToMeeting.NumberOfInvitedParticipants,0) as NumberOfInvitedParticipants
FROM
meeting meetingA
LEFT JOIN
(
   SELECT COUNT(*) as NumberOfInvitedParticipants, participantB.MeetingID
   FROM participant participantB
   INNER JOIN invitation ON participantB.InvitationID = invitation.InvitationID
   WHERE participantB.IsOrganizer = 0
         AND invitation.Requiredness = 1
   GROUP BY participantB.MeetingID
) InvitedToMeeting
ON meetingA.MeetingID = InvitedToMeeting.MeetingID
WHERE meetingA.MeetingID IN
        (
           SELECT participant.MeetingID
           FROM participant
           WHERE IsOrganizer = 1
                 AND participant.EmployeeID = @ParamEmployeeID
        )
        AND meetingA.StartTime >= @ParamStartTimeIntervalStart
        AND meetingA.StartTime <= @ParamStartTimeIntervalEnd
ORDER BY meetingA.StartTime
```

**Average number of required invited participants of meetings which were organized by a specified employee (where start time is in a specified interval)**

*@ParamEmployeeID = ID of the specified emplyoee*
*@ParamStartTimeIntervalStart = start time of the specified „start time" interval*
*@ParamStartTimeIntervalEnd = end time of the specified „start time" interval*

```
SELECT AVG(InvitedToMeeting.NumberOfInvitedParticipants) as AverageNumberOfInvitedParticipants
FROM
meeting meetingA
INNER JOIN
(
   SELECT COUNT(*) as NumberOfInvitedParticipants, participantB.MeetingID
   FROM participant participantB
   INNER JOIN invitation ON participantB.InvitationID = invitation.InvitationID
   WHERE participantB.IsOrganizer = 0
         AND invitation.Requiredness = 1
   GROUP BY participantB.MeetingID
) InvitedToMeeting
ON meetingA.MeetingID = InvitedToMeeting.MeetingID
WHERE meetingA.MeetingID IN
        (
           SELECT participant.MeetingID
           FROM participant
           WHERE IsOrganizer = 1
                 AND participant.EmployeeID = @ParamEmployeeID
        )
        AND meetingA.StartTime >= @ParamStartTimeIntervalStart
        AND meetingA.StartTime <= @ParamStartTimeIntervalEnd
GROUP BY NULL
```

## 5.15 Required versus optional participants

Percentage of required versus optional participants to be invited to a meeting.
Required participants may be more important (this is why they are required). If there are more optional participants than required, then these (supposedly less important) can interfere the meeting. A suggestion would be to invite less optional participants than required participants.

**Number of required and optional number of invited participants, who were invited to a specified meeting**

*@ParamMeetingID = ID of the specified meeting*

```
SELECT 'Required' as RequirednessType,
COALESCE((
   SELECT COUNT(*)
   FROM participant
   INNER JOIN invitation ON participant.InvitationID = invitation.InvitationID
   WHERE participant.MeetingID = @ParamMeetingID
   AND participant.IsOrganizer = 0
   AND invitation.Requiredness = 1
   GROUP BY participant.MeetingID
),0) as Value
UNION ALL
SELECT 'Optional' as RequirednessType,
COALESCE((
   SELECT COUNT(*)
   FROM participant
   INNER JOIN invitation ON participant.InvitationID = invitation.InvitationID
   WHERE participant.MeetingID = @ParamMeetingID
   AND participant.IsOrganizer = 0
   AND invitation.Requiredness = 0
   GROUP BY participant.MeetingID
),0) as Value
```

## 5.16 Time between meeting and core service hours

Minutes between meeting and core service hours. Depending on organisation culture and core service hoursare usually defined (e.g. from 8-10 to 15-17).
It may be risky to organize meetings before and after core service hours start/end or around other non-service hours (e.g. lunchtime). Some of the participants can decide not to join the meeting in these periods. Thus it could be useful to assure enough time between the meeting and core office hours. (e.g. start after 10 minutes of core meeting hours start, end 10 minutes before core office hours end) same principle can be applied to lunchtime. Naturally, meetings can be organized anytime if agreed with the participants, the above mentioned are rather suggestions than rules.

**Time between:**
 **- meeting start and core service hours start**
 **- meeting start and core service hours finish**
 **- meeting finish and core service hours start**

**- meeting finish and core service hours finish**
**In a specified meeting**

*@ParamMeetingID = ID of the specified meeting*
*@ParamCoreServiceStart= Core service hours start*
*@ParamCoreServiceFinish = Core service hours finish*

```
SELECT 'MeetingStart-CoreStart' as Type
COALESCE((
    SELECT TIMEDIFF(TIME(meeting.StartTime),@ParamCoreServiceStart)
    FROM meeting
    WHERE meeting.MeetingID = @ParamMeetingID
),0) as Value
UNION ALL
SELECT 'CoreFinish-MeetingStart' as Type,
COALESCE((
    SELECT TIMEDIFF(@ParamCoreServiceFinish ,TIME(meeting.StartTime))
    FROM meeting
    WHERE meeting.MeetingID = @ParamMeetingID
),0) as Value
UNION ALL
SELECT 'MeetingFinish-CoreStart' as Type,
COALESCE((
    SELECT TIMEDIFF(TIME(meeting.FinishTime),@ParamCoreServiceStart)
    FROM meeting
    WHERE meeting.MeetingID = @ParamMeetingID
),0) as Value
UNION ALL
SELECT 'CoreFinish-MeetingFinish' as Type,
COALESCE((
    SELECT TIMEDIFF(@ParamCoreServiceFinish ,TIME(meeting.FinishTime))
    FROM meeting
    WHERE meeting.MeetingID = @ParamMeetingID
),0) as Value
```

## 5.17   Time between meetings

Minutes between two consecutive meetings.

People need breaks. In contrast to this, some managers have meetings all day long. Sometimes this could be so exhausting that some people will not have time to relax or to prepare for meetings.
A calendar software could suggest best meeting time from an interval, taking into the account the calendars of all invitees assuring enough time for breaks and/or preparation.

**Time between two specified meeting**

Time between first meeting's finishtime and the second meeting's startime. If the second meeting starts before the first finishes, the result is 0.

*@ParamMeetingID1 = ID of the first specified meeting*
*@ParamMeetingID2 = ID of the second specified meeting*

```
SELECT
 GREATEST(
  GREATEST(TIMESTAMPDIFF(MINUTE,
    (SELECT FinishTime FROM meeting WHERE MeetingID = @ParamMeetingID1),
    (SELECT StartTime FROM meeting WHERE MeetingID = @ParamMeetingID2)
  ),0),
```

```
    GREATEST(TIMESTAMPDIFF(MINUTE,
      (SELECT FinishTime FROM meeting WHERE MeetingID = @ParamMeetingID2),
      (SELECT StartTime FROM meeting WHERE MeetingID = @ParamMeetingID1)
    ),0)) as Resul
```

## 5.18 Acceptive/tentative/non-responding/decline participants

Number of all participants who accepted/tentatively accepted/did not respond/declined the meeting request. This can be broken down to further metrics such as: number of required/optional participants who accepted the meeting request.

**Number of acceptive/tentative/non-responding/decline number of invited participants, who were invited to a specified meeting**

*@ParamMeetingID = ID of the specified meeting*

```
SELECT 'Accept' as Response,
COALESCE((
   SELECT COUNT(*)
   FROM participant
   INNER JOIN invitation ON participant.InvitationID = invitation.InvitationID
   INNER JOIN invitationresponse ON invitation.InvitationResponseID = invitationresponse.InvitationResponseID
   INNER JOIN responsetype ON invitationresponse.ResponseTypeID = responsetype.ResponseTypeID
   WHERE participant.MeetingID = @ParamMeetingID
   AND participant.IsOrganizer = 0
   AND responsetype.Name = 'accept'
   GROUP BY participant.MeetingID
),0) as Value
UNION ALL
SELECT 'Decline' as Response,
COALESCE((
   SELECT COUNT(*)
   FROM participant
   INNER JOIN invitation ON participant.InvitationID = invitation.InvitationID
   INNER JOIN invitationresponse ON invitation.InvitationResponseID = invitationresponse.InvitationResponseID
   INNER JOIN responsetype ON invitationresponse.ResponseTypeID = responsetype.ResponseTypeID
   WHERE participant.MeetingID = @ParamMeetingID
   AND participant.IsOrganizer = 0
   AND responsetype.Name = 'decline'
   GROUP BY participant.MeetingID
),0) as Value
UNION ALL
SELECT 'Non-Responding' as Response,
COALESCE((
   SELECT COUNT(*)
   FROM participant
   INNER JOIN invitation ON participant.InvitationID = invitation.InvitationID
   WHERE participant.MeetingID = @ParamMeetingID
   AND participant.IsOrganizer = 0
   AND ISNULL(invitation.InvitationResponseID)
   GROUP BY participant.MeetingID
),0) as Value
UNION ALL
SELECT 'Tentative' as Response,
COALESCE((
   SELECT COUNT(*)
   FROM participant
   INNER JOIN invitation ON participant.InvitationID = invitation.InvitationID
```

```
    INNER JOIN invitationresponse ON invitation.InvitationResponseID =
invitationresponse.InvitationResponseID
    INNER JOIN responsetype ON invitationresponse.ResponseTypeID =
responsetype.ResponseTypeID
    WHERE participant.MeetingID = @ParamMeetingID
    AND participant.IsOrganizer = 0
    AND responsetype.Name = 'tentative'
    GROUP BY participant.MeetingID
),0) as Value
```

## 5.19  Number of re-plan requests per meeting

Number of re-plan requests or number of times new time proposed. A meeting can have multiple re-plan requests e.g. if there is a conflict between the meeting and participants availability. Re-plan requests can raise awareness on discrepancies between the meeting interval and invitees availability. This could be due to location/other meetings already planned/meetings or tasks planned but not present in the invitees calendar.

Number of re-plan requests can be calculated (1) per meeting per person, (2) summed per person for a period of time (e.g. weekly/monthly) or (3) summed per meeting. In case if (1) is higher than 1, it could signal serious conflict between schedules of the chair and invitee. In case (2) is relatively high to other persons, it could signal that the attitude or workload of the person is relatively higher to others.

**List of invited participants with number of re-plan request, who were invited to a specified meeting**

*@ParamMeetingID = ID of the specified meeting*

```
SELECT employee.name as InvitedParticipants, COUNT(helper.InvitationID) as
NumberOfReplanRequest
FROM employee
INNER JOIN participant ON employee.EmployeeID = participant.EmployeeID
INNER JOIN invitation ON participant.InvitationID = invitation.InvitationID
LEFT JOIN
(
    SELECT inviteerequest.InvitationID
    FROM inviteerequest
    INNER JOIN inviteerequesttype
    ON inviteerequest.InviteeRequestTypeID = inviteerequesttype.InviteeRequestTypeID
    WHERE inviteerequesttype.Name = 're-plan'
) as helper
ON invitation.InvitationID = helper.InvitationID
WHERE participant.MeetingID = @ParamMeetingID
AND participant.IsOrganizer=0
GROUP BY employee.EmployeeID
```

**Sum of re-plan request of a specified meeting**

*@ParamMeetingID = ID of the specified meeting*

```
SELECT COUNT(helper.InvitationID) as NumberOfReplanRequest
FROM participant
INNER JOIN invitation ON participant.InvitationID = invitation.InvitationID
LEFT JOIN
(
    SELECT inviteerequest.InvitationID
    FROM inviteerequest
    INNER JOIN inviteerequesttype
    ON inviteerequest.InviteeRequestTypeID = inviteerequesttype.InviteeRequestTypeID
    WHERE inviteerequesttype.Name = 're-plan'
```

```sql
) as helper
ON invitation.InvitationID = helper.InvitationID
WHERE participant.MeetingID = @ParamMeetingID
AND participant.IsOrganizer=0
GROUP BY NULL
```

**List of numbers of re-plan request of meetings which were organized by a specified employee (where start time is in a specified interval)**

*@ParamEmployeeID = ID of the specified emplyoee*
*@ParamStartTimeIntervalStart = start time of the specified „start time" interval*
*@ParamStartTimeIntervalEnd = end time of the specified „start time" interval*

```sql
SELECT meeting.MeetingID, COUNT(helper.InvitationID) as NumberOfReplanRequest
FROM meeting
LEFT JOIN participant ON meeting.MeetingID = participant.MeetingID
                    AND participant.IsOrganizer = 0
LEFT JOIN invitation ON participant.InvitationID = invitation.InvitationID
LEFT JOIN
(
   SELECT inviteerequest.InvitationID
   FROM inviteerequest
   INNER JOIN inviteerequesttype
   ON inviteerequest.InviteeRequestTypeID = inviteerequesttype.InviteeRequestTypeID
   WHERE inviteerequesttype.Name = 're-plan'
) as helper
ON invitation.InvitationID = helper.InvitationID
WHERE meeting.MeetingID IN
        (
            SELECT participant.MeetingID
            FROM participant
            WHERE IsOrganizer = 1
                AND participant.EmployeeID = @ParamEmployeeID
        )
        AND meeting.StartTime >= @ParamStartTimeIntervalStart
        AND meeting.StartTime <= @ParamStartTimeIntervalEnd
GROUP BY meeting.MeetingID
```

**Average number of re-plan request of meetings which were organized by a specified employee (where start time is in a specified interval)**

*@ParamEmployeeID = ID of the specified emplyoee*
*@ParamStartTimeIntervalStart = start time of the specified „start time" interval*
*@ParamStartTimeIntervalEnd = end time of the specified „start time" interval*

```sql
SELECT AVG(counthelper.NumberOfReplanRequest) as Average
FROM
(
SELECT meeting.MeetingID, COUNT(helper.InvitationID) as NumberOfReplanRequest
FROM meeting
LEFT JOIN participant ON meeting.MeetingID = participant.MeetingID
                    AND participant.IsOrganizer = 0
LEFT JOIN invitation ON participant.InvitationID = invitation.InvitationID
LEFT JOIN
(
   SELECT inviteerequest.InvitationID
   FROM inviteerequest
   INNER JOIN inviteerequesttype
   ON inviteerequest.InviteeRequestTypeID = inviteerequesttype.InviteeRequestTypeID
   WHERE inviteerequesttype.Name = 're-plan'
) as helper
ON invitation.InvitationID = helper.InvitationID
WHERE meeting.MeetingID IN
        (
            SELECT participant.MeetingID
            FROM participant
```

```
            WHERE IsOrganizer = 1
                AND participant.EmployeeID = @ParamEmployeeID
        )
        AND meeting.StartTime >= @ParamStartTimeIntervalStart
        AND meeting.StartTime <= @ParamStartTimeIntervalEnd
GROUP BY meeting.MeetingID
) as counthelper
GROUP BY NULL
```

## 5.20 Number of information request participants

Number of invitees who requested more information. Some calendar software provide information request facility. Of course information request can happen in multiple forms e.g. face to face, in email, by phone or by chat. The higher this number probably the weaker the meeting invitation description is.

**List of invited participants with number of info request, who were invited to a specified meeting**

*@ParamMeetingID = ID of the specified meeting*

```
SELECT employee.name as InvitedParticipants, COUNT(helper.InvitationID) as
NumberOfReplanRequest
FROM employee
INNER JOIN participant ON employee.EmployeeID = participant.EmployeeID
INNER JOIN invitation ON participant.InvitationID = invitation.InvitationID
LEFT JOIN
(
    SELECT inviteerequest.InvitationID
    FROM inviteerequest
    INNER JOIN inviteerequesttype
    ON inviteerequest.InviteeRequestTypeID = inviteerequesttype.InviteeRequestTypeID
    WHERE inviteerequesttype.Name = 'info'
) as helper
ON invitation.InvitationID = helper.InvitationID
WHERE participant.MeetingID = @ParamMeetingID
AND participant.IsOrganizer=0
GROUP BY employee.EmployeeID
```

**Sum of info request of a specified meeting**

*@ParamMeetingID = ID of the specified meeting*

```
SELECT COUNT(helper.InvitationID) as NumberOfReplanRequest
FROM participant
INNER JOIN invitation ON participant.InvitationID = invitation.InvitationID
LEFT JOIN
(
    SELECT inviteerequest.InvitationID
    FROM inviteerequest
    INNER JOIN inviteerequesttype
    ON inviteerequest.InviteeRequestTypeID = inviteerequesttype.InviteeRequestTypeID
    WHERE inviteerequesttype.Name = 'info'
) as helper
ON invitation.InvitationID = helper.InvitationID
WHERE participant.MeetingID = @ParamMeetingID
AND participant.IsOrganizer=0
GROUP BY NULL
```

**List of numbers of info request of meetings which were organized by a specified employee (where start time is in a specified interval)**

*@ParamEmployeeID = ID of the specified emplyoee*
*@ParamStartTimeIntervalStart = start time of the specified „start time" interval*
*@ParamStartTimeIntervalEnd = end time of the specified „start time" interval*

```
SELECT meeting.MeetingID, COUNT(helper.InvitationID) as NumberOfReplanRequest
FROM meeting
LEFT JOIN participant ON meeting.MeetingID = participant.MeetingID
                    AND participant.IsOrganizer = 0
LEFT JOIN invitation ON participant.InvitationID = invitation.InvitationID
LEFT JOIN
(
   SELECT inviteerequest.InvitationID
   FROM inviteerequest
   INNER JOIN inviteerequesttype
   ON inviteerequest.InviteeRequestTypeID = inviteerequesttype.InviteeRequestTypeID
   WHERE inviteerequesttype.Name = 'info'
) as helper
ON invitation.InvitationID = helper.InvitationID
WHERE meeting.MeetingID IN
        (
            SELECT participant.MeetingID
            FROM participant
            WHERE IsOrganizer = 1
                AND participant.EmployeeID = @ParamEmployeeID
        )
        AND meeting.StartTime >= @ParamStartTimeIntervalStart
        AND meeting.StartTime <= @ParamStartTimeIntervalEnd
GROUP BY meeting.MeetingID
```

**Average number of info request of meetings which were organized by a specified employee (where start time is in a specified interval)**

*@ParamEmployeeID = ID of the specified emplyoee*
*@ParamStartTimeIntervalStart = start time of the specified „start time" interval*
*@ParamStartTimeIntervalEnd = end time of the specified „start time" interval*

```
SELECT AVG(counthelper.NumberOfReplanRequest) as Average
FROM
(
SELECT meeting.MeetingID, COUNT(helper.InvitationID) as NumberOfReplanRequest
FROM meeting
LEFT JOIN participant ON meeting.MeetingID = participant.MeetingID
                    AND participant.IsOrganizer = 0
LEFT JOIN invitation ON participant.InvitationID = invitation.InvitationID
LEFT JOIN
(
   SELECT inviteerequest.InvitationID
   FROM inviteerequest
   INNER JOIN inviteerequesttype
   ON inviteerequest.InviteeRequestTypeID = inviteerequesttype.InviteeRequestTypeID
   WHERE inviteerequesttype.Name = 'info'
) as helper
ON invitation.InvitationID = helper.InvitationID
WHERE meeting.MeetingID IN
        (
            SELECT participant.MeetingID
            FROM participant
            WHERE IsOrganizer = 1
                AND participant.EmployeeID = @ParamEmployeeID
        )
        AND meeting.StartTime >= @ParamStartTimeIntervalStart
        AND meeting.StartTime <= @ParamStartTimeIntervalEnd
GROUP BY meeting.MeetingID
) as counthelper
GROUP BY NULL
```

## 5.21 Meeting room changes

Number of meeting room changes. This metric could signal the uncertainty of the location. Some invitees may not receive the last change and could go to the wrong meeting room. This change is more expensive the more invitees are present and the distance between involved meeting rooms is higher and time between change notification and meeting start is lower.

**Number of meeting room changes of a specified meeting**

*@ParamMeetingID = ID of the specified meeting*

```
SELECT RoomChangedNumber as RoomChangedNumber
FROM meeting
WHERE meeting.MeetingID = @ParamMeetingID
```

*List of meetings' meeting room changes number which were organized by a specified employee (where start time is in a specified interval):*

*@ParamEmployeeID = ID of the specified emplyoee*
*@ParamStartTimeIntervalStart = start time of the specified „start time" interval*
*@ParamStartTimeIntervalEnd = end time of the specified „start time" interval*

```
SELECT meeting.MeetingID, RoomChangedNumber as RoomChangedNumber
FROM meeting
WHERE meeting.MeetingID IN
        (
          SELECT participant.MeetingID
          FROM participant
          WHERE IsOrganizer = 1
                AND participant.EmployeeID = @ParamEmployeeID
        )
        AND meeting.StartTime >= @ParamStartTimeIntervalStart
        AND meeting.StartTime <= @ParamStartTimeIntervalEnd
ORDER BY meeting.MeetingID ASC
```

**Average meeting room changes number of meetings which were organized by a specified employee (where start time is in a specified interval)**

*@ParamEmployeeID = ID of the specified emplyoee*
*@ParamStartTimeIntervalStart = start time of the specified „start time" interval*
*@ParamStartTimeIntervalEnd = end time of the specified „start time" interval*

```
SELECT AVG(RoomChangedNumber) as AverageRoomChangedNumber
FROM meeting
WHERE meeting.MeetingID IN
        (
          SELECT participant.MeetingID
          FROM participant
          WHERE IsOrganizer = 1
                AND participant.EmployeeID = @ParamEmployeeID
        )
        AND meeting.StartTime >= @ParamStartTimeIntervalStart
        AND meeting.StartTime <= @ParamStartTimeIntervalEnd
GROUP BY NULL
```

## 5.22 Held versus cancelled meetings

Number of held versus cancelled meetings. This metric can be measured per organisation, project, team or per employee. Cancelled meetings can cause re-planning of tasks of participants or can cause empty times in their schedule (due to the time needed for context switching or in case when no other tasks can be quickly scheduled for the freed period).

**A specified meeting is cancelled or not**

*@ParamMeetingID = ID of the specified meeting*

```
SELECT meeting.Cancelled
FROM meeting
WHERE meeting.MeetingID = @ParamMeetingID
```

**Number of held, and number of cancelled meetings which were organized by a specified employee (where start time is in a specified interval)**

*@ParamEmployeeID = ID of the specified emplyoee*
*@ParamStartTimeIntervalStart = start time of the specified „start time" interval*
*@ParamStartTimeIntervalEnd = end time of the specified „start time" intervalend time of the specified „start time" interval*

```
SELECT 'Held' as Type,
COALESCE((
   SELECT COUNT(*)
   FROM meeting
   WHERE
        meeting.MeetingID IN
        (
           SELECT participant.MeetingID
           FROM participant
           WHERE IsOrganizer = 1
                AND participant.EmployeeID = @ParamEmployeeID
        )
        AND meeting.StartTime >= @ParamStartTimeIntervalStart
        AND meeting.StartTime <= @ParamStartTimeIntervalEnd
        AND meeting.Cancelled = 0
   GROUP BY meeting.Cancelled
),0) as Value
UNION ALL
SELECT 'Cancelled' as Type,
COALESCE((
   SELECT COUNT(*)
   FROM meeting
   WHERE
        meeting.MeetingID IN
        (
           SELECT participant.MeetingID
           FROM participant
           WHERE IsOrganizer = 1
                AND participant.EmployeeID = @ParamEmployeeID
        )
        AND meeting.StartTime >= @ParamStartTimeIntervalStart
        AND meeting.StartTime <= @ParamStartTimeIntervalEnd
        AND meeting.Cancelled = 1
   GROUP BY meeting.Cancelled
),0) as Value
```

## 5.23 Number of consecutive meetings

Number of meetings making one block. It can be calculated (1) per person or (2) per meeting rooms (e.g. for calculating the resource efficiency of the meeting room).

**Specified two meeting is consecutive or not (return 1 if consecutive, return 0 if not consecutive). Two meeting is consecutive if time between them lower then 10 minutes.**

*@ParamMeetingID1 = ID of the first specified meeting*
*@ParamMeetingID2 = ID of the second specified meeting*

```
SELECT CASE WHEN helper.Result < 10 THEN 1 ELSE 0 END as IsConsecutive
FROM (
SELECT
 GREATEST(
  GREATEST(TIMESTAMPDIFF(MINUTE,
     (SELECT FinishTime FROM meeting WHERE MeetingID = @ParamMeetingID1),
     (SELECT StartTime FROM meeting WHERE MeetingID = @ParamMeetingID2)
  ),0),
  GREATEST(TIMESTAMPDIFF(MINUTE,
     (SELECT FinishTime FROM meeting WHERE MeetingID = @ParamMeetingID2),
     (SELECT StartTime FROM meeting WHERE MeetingID = @ParamMeetingID1)
  ),0)) as Result
) as helper
```

**All consecutive meeting pairs in a specified employee's meetings (where start time is in a specified interval)**

*@ParamEmployeeID = ID of the specified emplyoee*
*@ParamStartTimeIntervalStart = start time of the specified „start time" interval*
*@ParamStartTimeIntervalEnd = end time of the specified „start time" intervalend time of the specified „start time" interval*

```
SELECT helper.ID1, helper.ID2, helper.TimeBetween
FROM
(
SELECT meetingA.MeetingID as ID1, meetingB.MeetingID as ID2,
  GREATEST(
     GREATEST(TIMESTAMPDIFF(MINUTE,meetingA.FinishTime,meetingB. StartTime),0),
     GREATEST(TIMESTAMPDIFF(MINUTE,meetingB.FinishTime,meetingA. StartTime),0)
  ) as TimeBetween
FROM
(meeting meetingA
INNER JOIN participant participantA ON participantA.MeetingID = meetingA.MeetingID)
CROSS JOIN
(meeting meetingB
INNER JOIN participant participantB ON participantB.MeetingID = meetingB.MeetingID)
WHERE meetingA.MeetingID < meetingB.MeetingID
        AND meetingA.StartTime >= @ParamStartTimeIntervalStart
        AND meetingA.StartTime <= @ParamStartTimeIntervalEnd
        AND meetingB.StartTime >= @ParamStartTimeIntervalStart
        AND meetingB.StartTime <= @ParamStartTimeIntervalEnd
        AND participantA.EmployeeID = @ParamEmployeeID
        AND participantB.EmployeeID = @ParamEmployeeID
        AND participantA.IsOrganizer = 1
        AND participantB.IsOrganizer = 1
) as helper
WHERE helper.TimeBetween < 10
```

## 5.24 Number of meetings

Number of meetings a person is invited/attends per day/week/month. This metric can be used for calculating human resource availability, load distribution, or can show employees with highest/least load. This can be a comparative metric if measurement is done on the same role.

**Number of meeting / person / specified interval**

*@ParamEmployeeID = ID of the specified emplyoee*
*@ParamStartTimeIntervalStart = start time of the specified „start time" interval*
*@ParamStartTimeIntervalEnd = end time of the specified „start time" intervalend time of the specified „start time" interval*

```
SELECT COUNT(*)
FROM participant
INNER JOIN meeting ON meeting.MeetingID = participant.MeetingID
WHERE participant.EmployeeID = @ParamEmployeeID
        AND meeting.StartTime >= @ParamStartTimeIntervalStart
        AND meeting.StartTime <= @ParamStartTimeIntervalEnd
GROUP BY participant.EmployeeID
```

## 5.25 Number of parallel meeting hours

This metric is a more precise variant of "Number or parallel meetings / person…" metric, providing a hourly view of meeting overlapping instead of the number of occurrences representation. It is identical to the number of occurrences in the case when the start and end of parallel meetings are occurring in the same time (e.g. in the same half an hour) or in the case when are fully overlapped (starting and ending in the same time). Both approaches have their value added: time (hours) based approach is more precise, providing a detailed picture, while occurrence based in on a higher abstraction level and thus can provide easily an overall picture.

**Count of parallel meeting pairs in a specified employee's meetings (where start time is in a specified interval)**

*@ParamEmployeeID = ID of the specified emplyoee*
*@ParamStartTimeIntervalStart = start time of the specified „start time" interval*
*@ParamStartTimeIntervalEnd = end time of the specified „start time" intervalend time of the specified „start time" interval*

```
SELECT SUM(helper2.TimeBetween) as NumberOfParallelMeetingPairs
FROM
(
SELECT helper.ID1, helper.ID2, helper.TimeBetween
FROM
(
SELECT meetingA.MeetingID as ID1, meetingB.MeetingID as ID2,
  TIMESTAMPDIFF(MINUTE,
    GREATEST(meetingA.StartTime,meetingB.StartTime),
    LEAST(meetingA.FinishTime,meetingB.FinishTime)
  ) as TimeBetween
FROM
(meeting meetingA
```

```
INNER JOIN participant participantA ON participantA.MeetingID = meetingA.MeetingID)
CROSS JOIN
(meeting meetingB
INNER JOIN participant participantB ON participantB.MeetingID = meetingB.MeetingID)
WHERE meetingA.MeetingID < meetingB.MeetingID
        AND meetingA.StartTime >= @ParamStartTimeIntervalStart
        AND meetingA.StartTime <= @ParamStartTimeIntervalEnd
        AND meetingB.StartTime >= @ParamStartTimeIntervalStart
        AND meetingB.StartTime <= @ParamStartTimeIntervalEnd
        AND participantA.EmployeeID = @ParamEmployeeID
        AND participantB.EmployeeID = @ParamEmployeeID
        AND participantA.IsOrganizer = 1
        AND participantB.IsOrganizer = 1
) as helper
WHERE helper.TimeBetween > 0
) as helper2
```

## 5.26 Chaired versus attended meetings among two or more persons

This is a variant of "Ratio of chaired versus attended meetings" metric and provides the dynamics between two or more employees. Depending on organization culture and meeting rules, it may serve as a comparative measure on proactiveness.

**Number of chaired and number of attended meeting / 2 person / specified interval**

*@ParamEmployeeID1 = ID of the specified emplyoee*
*@ParamEmployeeID2 = ID of the specified emplyoee*
*@ParamStartTimeIntervalStart = start time of the specified „start time" interval*
*@ParamStartTimeIntervalEnd = end time of the specified „start time" interval*

```
SELECT participant.EmployeeID, participant.IsOrganizer, COUNT(*) as CountParticipant
FROM participant
INNER JOIN meeting ON meeting.MeetingID = participant.MeetingID
WHERE (participant.EmployeeID = @ParamEmployeeID1 OR participant.EmployeeID =
@ParamEmployeeID2)
        AND meeting.StartTime >= @ParamStartTimeIntervalStart
        AND meeting.StartTime <= @ParamStartTimeIntervalEnd
GROUP BY participant.EmployeeID, participant.IsOrganizer
```

## 5.27 Number of average participants on a meeting that a person chairs

This metric provides the average number of participants of a meeting organized by a person. Persons organizing meetings with high number of participants in average may enjoy public speaking and could have a key role in the organization (e.g. leaders or trainers).

Organizing meetings with high number of participants might be affected by some other meetings with a lower number of participants. Since meetings with less participants can be reorganized more flexibly a practice could be to organize larger meetings/events even if they overlap with smaller meetings.

**Number of average participants on a meeting that a specified person chairs (in a specified time interval)**

*@ParamEmployeeID = ID of the specified emplyoee*
*@ParamStartTimeIntervalStart = start time of the specified „start time" interval*
*@ParamStartTimeIntervalEnd = end time of the specified „start time" intervalend time of the specified „start time" interval*

```
SELECT IFNULL(AVG(helper.CountParticipant),0) as Average
FROM
(
SELECT meeting.MeetingID as MeetingID, COUNT(*) as CountParticipant
FROM participant
INNER JOIN meeting ON meeting.MeetingID = participant.MeetingID
WHERE
        participant.MeetingID IN
        (
           SELECT participant.MeetingID
           FROM participant
           WHERE IsOrganizer = 1
                 AND participant.EmployeeID = @ParamEmployeeID
        )
        AND meeting.StartTime >= @ParamStartTimeIntervalStart
        AND meeting.StartTime <= @ParamStartTimeIntervalEnd
GROUP BY meeting.MeetingID
) as helper
```

## 5.28  Most frequent participants on meetings that a person chairs

This metric provides list of most frequent participants on a meeting. Shows those people which are most connected with the chair. List can be compared with organization chart/job descriptions and could highlight organizational inefficiencies and opportunities to reposition/reorganize teams.

**Number of most frequent participants on a meeting that a specified person chairs (in a specified time interval)**

*@ParamEmployeeID = ID of the specified emplyoee*
*@ParamStartTimeIntervalStart = start time of the specified „start time" interval*
*@ParamStartTimeIntervalEnd = end time of the specified „start time" interval*

```
SELECT helper.CountParticipant, COUNT(*) as CountParticipantFrequency
FROM
(
SELECT meeting.MeetingID as MeetingID, COUNT(*) as CountParticipant
FROM participant
INNER JOIN meeting ON meeting.MeetingID = participant.MeetingID
WHERE
        participant.MeetingID IN
        (
           SELECT participant.MeetingID
           FROM participant
           WHERE IsOrganizer = 1
                 AND participant.EmployeeID = @ParamEmployeeID
        )
        AND meeting.StartTime >= @ParamStartTimeIntervalStart
        AND meeting.StartTime <= @ParamStartTimeIntervalEnd
GROUP BY meeting.MeetingID
) as helper
GROUP BY helper.CountParticipant
ORDER BY CountParticipantFrequency DESC
```

## 5.29 Average number of meetings

This metric provides average number of meetings of a person in a daily, weekly or monthly breakdown. This metric can be used as a load indicator. The load indicator could be used for identifying capacity problems e.g. the need of load redistribution among people having the same role or the need of new hire for satisfying resource needs.

**Number of average meeting / person / day (in a specified time interval)**

*@ParamEmployeeID = ID of the specified emplyoee*
*@ParamStartTimeIntervalStart = start time of the specified „start time" interval*
*@ParamStartTimeIntervalEnd = end time of the specified „start time" interval*

```
SELECT AVG(helper.Meetings) as Result
FROM
(
SELECT COUNT(*) as Meetings, DATE(meeting.StartTime) as Time
FROM participant
INNER JOIN meeting ON meeting.MeetingID = participant.MeetingID
WHERE participant.EmployeeID = @ParamEmployeeID
      AND meeting.StartTime >= @ParamStartTimeIntervalStart
      AND meeting.StartTime <= @ParamStartTimeIntervalEnd
GROUP BY DATE(meeting.StartTime)
) as helper
```

**Number of average meeting / person / week (in a specified time interval)**

*@ParamEmployeeID = ID of the specified emplyoee*
*@ParamStartTimeIntervalStart = start time of the specified „start time" interval*
*@ParamStartTimeIntervalEnd = end time of the specified „start time" interval*

```
SELECT AVG(helper.Meetings) as Result
FROM
(
SELECT COUNT(*) as Meetings, YEARWEEK(meeting.StartTime) as Time
FROM participant
INNER JOIN meeting ON meeting.MeetingID = participant.MeetingID
WHERE participant.EmployeeID = @ParamEmployeeID
      AND meeting.StartTime >= @ParamStartTimeIntervalStart
      AND meeting.StartTime <= @ParamStartTimeIntervalEnd
GROUP BY YEARWEEK(meeting.StartTime)
) as helper
```

**Number of average meeting / person / month (in a specified time interval)**

*@ParamEmployeeID = ID of the specified emplyoee*
*@ParamStartTimeIntervalStart = start time of the specified „start time" interval*
*@ParamStartTimeIntervalEnd = end time of the specified „start time" interval*

```
SELECT AVG(helper.Meetings) as Result
FROM
(
SELECT COUNT(*) as Meetings, YEAR(meeting.StartTime), MONTH(meeting.StartTime)  as Time
FROM participant
INNER JOIN meeting ON meeting.MeetingID = participant.MeetingID
WHERE participant.EmployeeID = @ParamEmployeeID
      AND meeting.StartTime >= @ParamStartTimeIntervalStart
      AND meeting.StartTime <= @ParamStartTimeIntervalEnd
GROUP BY YEAR(meeting.StartTime), MONTH(meeting.StartTime)
) as helper
```

## 5.30 Ratio of booked versus available meeting rooms

This metric is an indicator for meeting room availability. Ratio of booked and free meeting rooms can calculated per multiple timeframes (e.g. per hour).
The shorter the period the metric is calculated the finer the grain of information is. A quarter-hourly data collection will show optimization options for partially overlapping meetings.
Meeting peak times can also be collected and additional meeting rooms could be allocated dynamically for peak periods in a rented/co-working office by renting more rooms only for peak periods.

**At a specified datetime, a specified meeting room is available, or not (if the result is 0, then the room is available, if the result is >0 then the room is booked).**

*@ParamDateTime = the specified datetime*
*@ParamRoomID = the specified meeting room*

```
SELECT Count(*) as Result
FROM
(
  SELECT location.RoomID
  FROM location
  INNER JOIN locationinmeeting ON locationinmeeting.LocationID = location.LocationID
  INNER JOIN meeting ON locationinmeeting.MeetingID = meeting.MeetingID
  WHERE @ParamDateTime >= meeting.StartTime
        AND @ParamDateTime <= meeting.FinishTime
        AND NOT ISNULL(location.RoomID)
) as BookedRooms
WHERE BookedRooms.RoomID = @ParamRoomID
```

**At a specified datetime, which meeting rooms are available**

*@ParamDateTime = the specified datetime*

```
SELECT RoomID
FROM room
WHERE room.RoomID NOT IN
(
  SELECT location.RoomID
  FROM location
  INNER JOIN locationinmeeting ON locationinmeeting.LocationID = location.LocationID
  INNER JOIN meeting ON locationinmeeting.MeetingID = meeting.MeetingID
  WHERE @ParamDateTime >= meeting.StartTime
        AND @ParamDateTime <= meeting.FinishTime
        AND NOT ISNULL(location.RoomID)
)
```

## 5.31 Most coupled employee

The person having meetings with the highest number of other persons.
**Most coupled employee in a specified time interval**

*@ParamStartTimeIntervalStart = start time of the specified „start time" interval*
*@ParamStartTimeIntervalEnd = end time of the specified „start time" interval*

```
SELECT employee.Name, SUM(helper.number) as number
FROM
```

```
(
SELECT participantA.EmployeeID as Par1, participantB.EmployeeID as Par2, COUNT(*) as number
FROM participant participantA
INNER JOIN participant participantB ON participantA.MeetingID = participantB.MeetingID
INNER JOIN meeting ON meeting.MeetingID = participantA.MeetingID
WHERE participantA.EmployeeID <> participantB.EmployeeID
        AND meeting.StartTime >= @ParamStartTimeIntervalStart
        AND meeting.StartTime <= @ParamStartTimeIntervalEnd
GROUP BY participantA.EmployeeID, participantB.EmployeeID
HAVING number > ParamLowerLimit
) as helper
INNER JOIN employee ON helper.Par1 = employee.EmployeeID
GROUP BY helper.Par1
ORDER BY number DESC
```

## 5.32 Most coupled employee with lower limit

This metric is similar to the "Most coupled employee" metric, however in this case only those employees are considered with which the number of meetings reaches a lower limit. Introducing the lower limit will show employees who frequently/usually have meetings with many other employees.

**Most coupled employee with lower limit in a specified time interval**

*@ParamStartTimeIntervalStart = start time of the specified „start time" interval*
*@ParamStartTimeIntervalEnd = end time of the specified „start time" interval*
*@ParamLowerLimit = the lower limit*

```
SELECT employee.Name, SUM(helper.number) as number
FROM
(
SELECT participantA.EmployeeID as Par1, participantB.EmployeeID as Par2, COUNT(*) as number
FROM participant participantA
INNER JOIN participant participantB ON participantA.MeetingID = participantB.MeetingID
INNER JOIN meeting ON meeting.MeetingID = participantA.MeetingID
WHERE participantA.EmployeeID <> participantB.EmployeeID
        AND meeting.StartTime >= @ParamStartTimeIntervalStart
        AND meeting.StartTime <= @ParamStartTimeIntervalEnd
GROUP BY participantA.EmployeeID, participantB.EmployeeID
HAVING number > ParamLowerLimit
) as helper
INNER JOIN employee ON helper.Par1 = employee.EmployeeID
GROUP BY helper.Par1
ORDER BY number DESC
```

## 5.33 Most coupled employee with upper limit

This metric is similar to the "Most coupled employee" metric, however in this case only those employees are considered with which the number of meetings reaches an upper limit. Introducing the upper limit will show employees who have meetings with many other employees for a short period of time. Leaders and trainers may belong to this set (having a small number of meetings, but with many employees – e.g. when they train new employees).

**Most coupled employee with upper limit in a specified time interval**

*@ParamStartTimeIntervalStart = start time of the specified „start time" interval*
*@ParamStartTimeIntervalEnd = end time of the specified „start time" interval*
*@ParamUpperLimit = the upper limit*

```sql
SELECT employee.Name, SUM(helper.number) as number
FROM
(
SELECT participantA.EmployeeID as Par1, participantB.EmployeeID as Par2, COUNT(*) as number
FROM participant participantA
INNER JOIN participant participantB ON participantA.MeetingID = participantB.MeetingID
INNER JOIN meeting ON meeting.MeetingID = participantA.MeetingID
WHERE participantA.EmployeeID <> participantB.EmployeeID
        AND meeting.StartTime >= @ParamStartTimeIntervalStart
        AND meeting.StartTime <= @ParamStartTimeIntervalEnd
GROUP BY participantA.EmployeeID, participantB.EmployeeID
HAVING number < ParamUpperLimit
) as helper
INNER JOIN employee ON helper.Par1 = employee.EmployeeID
GROUP BY helper.Par1
ORDER BY number DESC
```

## 5.34 Number of participants, number of non-invited participants, active/inactive participants

This metric provides an overview on the composition of participants: number of invited and number of non-invited (but present) participants. Number of invited participants can be decomposed into two further sets: present and not present on the meeting. An ideal case is when the total number of participants is equals with invited and present participants. If the number non-invited participants is high it could signal that invitations are propagated informally or indirectly. In case if many or most important invitees are not present, the meeting might not achieve its goal and might be time consuming for those present.

**Number of „invited and active", „invited and inactive", and „non-invited but active" participants in a specified meeting**

*@ParamMeetingID = ID of the specified meeting*

```sql
SELECT 'InvitedAndActive' as Response,
COALESCE((
   SELECT COUNT(*)
   FROM participant
   INNER JOIN invitation ON participant.InvitationID = invitation.InvitationID
   WHERE participant.MeetingID = @ParamMeetingID
   AND participant.IsOrganizer = 0
   AND participant.ParticipateOnMeeting = 1
   GROUP BY participant.MeetingID
),0) as Value
UNION ALL
SELECT 'InvitedAndInActive' as Response,
COALESCE((
   SELECT COUNT(*)
   FROM participant
   INNER JOIN invitation ON participant.InvitationID = invitation.InvitationID
   WHERE participant.MeetingID = @ParamMeetingID
   AND participant.IsOrganizer = 0
```

```
    AND participant.ParticipateOnMeeting = 0
    GROUP BY participant.MeetingID
),0) as Value
UNION ALL
SELECT 'NonInvitedButActive' as Response,
COALESCE((
    SELECT COUNT(*)
    FROM participant
    WHERE participant.MeetingID = @ParamMeetingID
    AND participant.IsOrganizer = 0
    AND participant.ParticipateOnMeeting = 1
    AND ISNULL(participant.InvitationID)
    GROUP BY participant.MeetingID
),0) as Value
```

## 5.35 Most chaired employees

Top list of employees organizing the most meetings in the organization. This list in one hand could highlight people with good organization/communication skills, on the other hand can be used for notifying "heavy" meeting organizers in cases when there is a limited number of meeting rooms to step down and let others organize their meetings too.

**Most chaired employee in a specified time interval**

*@ParamStartTimeIntervalStart = start time of the specified „start time" interval*
*@ParamStartTimeIntervalEnd = end time of the specified „start time" interval*

```
SELECT EmployeeID, COUNT(*) as number
FROM participant
INNER JOIN employee ON employee.EmployeeID = participant.EmployeeID
INNER JOIN meeting ON meeting.MeetingID = participant.MeetingID
WHERE IsOrganizer = 1
        AND meeting.StartTime >= @ParamStartTimeIntervalStart
        AND meeting.StartTime <= @ParamStartTimeIntervalEnd
GROUP BY EmployeeID
ORDER BY number DESC
LIMIT 1
```

## 5.36 Most overloaded employees

Most overloaded employees can be shown in multiple ways. One method is generating a list of persons with top parallel meetings in a time period (e.g. list of people with highest number of overlapping meetings in a month). Another way of finding overloaded employees is to list those who have the highest number of meetings in a time period.

**List of employees and their meeting numbers in specified time interval. The list is ordered by the meeting numbers, first is the highest.**

*@ParamStartTimeIntervalStart = start time of the specified „start time" interval*
*@ParamStartTimeIntervalEnd = end time of the specified „start time" interval*

```
SELECT participant.EmployeeID, COUNT(*) as NumberOfMeetings
FROM participant
INNER JOIN meeting ON meeting.MeetingID = participant.MeetingID
WHERE meeting.StartTime >= @ParamStartTimeIntervalStart
```

```
            AND meeting.StartTime <= @ParamStartTimeIntervalEnd
GROUP BY participant. EmployeeID
ORDER BY NumberOfMeetings DESC
```

**List of employees (who has parallel meeting hours) and their total parallel meeting hours in specified time interval.**

*@ParamStartTimeIntervalStart = start time of the specified „start time" interval*
*@ParamStartTimeIntervalEnd = end time of the specified „start time" intervalend time of the specified „start time" interval*

```
SELECT helper2.Emp, SUM(helper2.TimeBetween) as NumberOfParallelMeetingPairs
FROM
(
   SELECT helper.ID1, helper.ID2, helper.TimeBetween, helper.Emp
   FROM
   (
      SELECT meetingA.MeetingID as ID1, meetingB.MeetingID as ID2, participantA.EmployeeID as Emp,
        TIMESTAMPDIFF(MINUTE,
         GREATEST(meetingA.StartTime,meetingB.StartTime),
         LEAST(meetingA.FinishTime,meetingB.FinishTime)
         ) as TimeBetween
      FROM
      (meeting meetingA
         INNER JOIN participant participantA ON participantA.MeetingID = meetingA.MeetingID)
      CROSS JOIN
      (meeting meetingB
         INNER JOIN participant participantB ON participantB.MeetingID = meetingB.MeetingID)
      WHERE meetingA.MeetingID < meetingB.MeetingID
         AND meetingA.StartTime >= @ParamStartTimeIntervalStart
         AND meetingA.StartTime <= @ParamStartTimeIntervalEnd
         AND meetingB.StartTime >= @ParamStartTimeIntervalStart
         AND meetingB.StartTime <= @ParamStartTimeIntervalEnd
         AND participantA.EmployeeID = participantB.EmployeeID
   ) as helper
   WHERE helper.TimeBetween > 0
) as helper2
GROUP BY helper2.Emp
```

## 5.37  Most tentative employees

Top list of employees, who provided most of the tentative answers to meeting invitations. These employees might reduce predictability and planning and thus efficiency of the meeting organization process.

**Most tentative employees in a specified time interval**

*@ParamStartTimeIntervalStart = start time of the specified „start time" interval*
*@ParamStartTimeIntervalEnd = end time of the specified „start time" interval*

```
SELECT EmployeeID, COUNT(*) as number
FROM participant
INNER JOIN invitation ON participant.InvitationID = invitation.InvitationID
INNER JOIN invitationresponse ON invitation.InvitationResponseID = invitationresponse.InvitationResponseID
INNER JOIN responsetype ON invitationresponse.ResponseTypeID = responsetype.ResponseTypeID
WHERE participant.IsOrganizer = 0 AND responsetype.Name = 'tentative'
GROUP BY EmployeeID
ORDER BY number DESC
```

## 5.38 Average time between meeting invitation and meeting start

Average time between organizing (sending out the meeting request) and meeting start per employees per timeframe. This metric can be used by individuals in refining their meeting organization strategy. Meeting requests sent out too early may signal that the meeting is not urgent, therefore some of the invitees may forget it or may consider it unimportant. In addition, it is easier to plan for a shorter period (e.g. for a couple of days or week than for a month or year). Meetings planned couple of weeks in advance will carry the risk of re-planning or to be replaced by incoming, more urgent tasks. On the other hand, very short average times between meeting invitations and meetings (e.g. couple of hours) could end in inefficiency with no time for preparation.

**Average time between meeting invitation and meeting start where a specified person chaired the meeting (and the start time is in a specified time interval)**

*@ ParamEmployeeID = ID of the specified emplyoee*
*@ParamStartTimeIntervalStart = start time of the specified „start time" interval*
*@ParamStartTimeIntervalEnd = end time of the specified „start time" interval*

```
SELECT SEC_TO_TIME(AVG(TIME_TO_SEC(TIMEDIFF(meeting.StartTime,meeting.InvitationTime))))
as AverageTime
FROM meeting
WHERE meeting.MeetingID IN
        (
            SELECT participant.MeetingID
            FROM participant
            WHERE IsOrganizer = 1
                AND participant.EmployeeID = @ParamEmployeeID
        )
        AND meeting.StartTime >= @ParamStartTimeIntervalStart
        AND meeting.StartTime <= @ParamStartTimeIntervalEnd
GROUP BY NULL
```

## 5.39 Lowest/Highest meeting cancellation

Highest and lowest number of meetings cancelled per person per timeframe. Employees who cancel meetings can slow down the organizational meeting organization process. Let's imagine that employee A has booked the last available meeting room for a given hour, then employees B-Z are wishing to book a room. They have multiple options, some examples: (1) will not organize the meeting, because there are no free meeting rooms, (2) will postpone the meeting to another timeframe, (3) or will recheck multiple times if a meeting room became free. It is likely that employees B-Z may spend additional time on booking their meetings when rooms are booked and cancelled. Therefore low cancellation rate is preferred for all employees. Employees with high cancellation rates may be notified to refine their meeting organization habits.

**Most cancellation by employees in a specified time interval**

*@ParamStartTimeIntervalStart = start time of the specified „start time" interval*
*@ParamStartTimeIntervalEnd = end time of the specified „start time" interval*

```
SELECT EmployeeID, COUNT(*) as number
FROM meeting
INNER JOIN participant ON participant.MeetingID = meeting.MeetingID
WHERE IsOrganizer = 1
        AND meeting.StartTime >= @ParamStartTimeIntervalStart
        AND meeting.StartTime <= @ParamStartTimeIntervalEnd
        AND Cancelled = 1
GROUP BY EmployeeID
ORDER BY number DESC
```

## 5.40  Loosely coupled employees

Top list of employees in a role with lowest number of meetings in a timeframe (e.g. in a year). This can be compared with an official organisation chart and could serve as a basis of identifying discrepancies in roles (e.g. managers/leaders might not be loosely coupled). For other roles, where meetings are more rare (e.g. software developers) this metric would not signal that they are loosely coupled. Number of meetings above average for a software developer could signal that s(he) could be a good performer in a management role too.

**List of employees and their meeting numbers in specified time interval. The list is ordered by the meeting numbers, first is the lowest.**

*@ParamStartTimeIntervalStart = start time of the specified „start time" interval*
*@ParamStartTimeIntervalEnd = end time of the specified „start time" interval*

```
SELECT participant.EmployeeID, COUNT(*) as NumberOfMeetings
FROM participant
INNER JOIN meeting ON meeting.MeetingID = participant.MeetingID
WHERE meeting.StartTime >= @ParamStartTimeIntervalStart
        AND meeting.StartTime <= @ParamStartTimeIntervalEnd
GROUP BY participant. EmployeeID
ORDER BY NumberOfMeetings ASC
```

## 5.41  Delegated meetings per person

This metric provides the percentage of meetings delegated to other persons. A high value obtained for this metric (e.g. 50% of meetings delegated) could be interpreted in many different ways. Some examples: it could signal that the person is good at delegation and may have other management skills, or s(he) is overloaded and may need some external help. Additionally, in case if this number is high it may reduce plannability of the work of those to whom meetings are delegated to.

**Number of delegated meetings / person / specified interval**

*@ParamEmployeeID = ID of the specified emplyoee*
*@ParamStartTimeIntervalStart = start time of the specified „start time" interval*
*@ParamStartTimeIntervalEnd = end time of the specified „start time" interval*

```
SELECT COUNT(*) as NumberOfDelegatedMeetings
FROM participant
INNER JOIN meeting ON meeting.MeetingID = participant.MeetingID
```

```
INNER JOIN invitation ON invitation.InvitationID = participant.InvitationID
WHERE participant.EmployeeID = @ParamEmployeeID
        AND meeting.StartTime >= @ParamStartTimeIntervalStart
        AND meeting.StartTime <= @ParamStartTimeIntervalEnd
        AND NOT ISNULL(invitation.DelegatedTo)
GROUP BY participant. EmployeeID
```

## 5.42  Received delegation requests

Delegation requests received in a period of time. This metric can be analysed as a whole, counting all numbers of delegation requests within a company, or per persons. People receiving most delegation requests may be trusted employees by their supervisors or colleagues.

**Number of received delegation requests / person / specified interval**

*@ParamEmployeeID = ID of the specified emplyoee*
*@ParamStartTimeIntervalStart = start time of the specified „start time" interval*
*@ParamStartTimeIntervalEnd = end time of the specified „start time" interval*

```
SELECT COUNT(*) as NumberOfDelegatedMeetings
FROM participant
INNER JOIN meeting ON meeting.MeetingID = participant.MeetingID
INNER JOIN invitation ON invitation.InvitationID = participant.InvitationID
WHERE participant.EmployeeID = @ParamEmployeeID
        AND meeting.StartTime >= @ParamStartTimeIntervalStart
        AND meeting.StartTime <= @ParamStartTimeIntervalEnd
        AND NOT ISNULL(invitation.DelegatedFrom)
GROUP BY participant. EmployeeID
```

## 5.43  Accepted versus non-replied versus rejected delegation requests

The percentage of accepted versus non-replied versus rejected delegation requests can be calculated per person or per period. When calculated per person it could be calculated both at least three different roles: sender, receiver and meeting organizer persons.

**Number of acceptive/tentative/non-responding/decline number of participants, who were delegated to a specified meeting**

*@ParamMeetingID = ID of the specified meeting*

```
SELECT 'Accept' as Response,
COALESCE((
   SELECT COUNT(*)
   FROM participant
   INNER JOIN invitation ON participant.InvitationID = invitation.InvitationID
   INNER JOIN invitationresponse ON invitation.InvitationResponseID =
invitationresponse.InvitationResponseID
   INNER JOIN responsetype ON invitationresponse.ResponseTypeID =
responsetype.ResponseTypeID
   WHERE participant.MeetingID = @ParamMeetingID
   AND participant.IsOrganizer = 0
   AND responsetype.Name = 'accept'
   AND NOT ISNULL(invitation.DelegatedFrom)
   GROUP BY participant.MeetingID
),0) as Value
UNION ALL
```

```sql
SELECT 'Decline' as Response,
COALESCE((
    SELECT COUNT(*)
    FROM participant
    INNER JOIN invitation ON participant.InvitationID = invitation.InvitationID
    INNER JOIN invitationresponse ON invitation.InvitationResponseID =
invitationresponse.InvitationResponseID
    INNER JOIN responsetype ON invitationresponse.ResponseTypeID =
responsetype.ResponseTypeID
    WHERE participant.MeetingID = @ParamMeetingID
    AND participant.IsOrganizer = 0
    AND responsetype.Name = 'decline'
    AND NOT ISNULL(invitation.DelegatedFrom)
    GROUP BY participant.MeetingID
),0) as Value
UNION ALL
SELECT 'Non-Responding' as Response,
COALESCE((
    SELECT COUNT(*)
    FROM participant
    INNER JOIN invitation ON participant.InvitationID = invitation.InvitationID
    WHERE participant.MeetingID = @ParamMeetingID
    AND participant.IsOrganizer = 0
    AND ISNULL(invitation.InvitationResponseID)
    AND NOT ISNULL(invitation.DelegatedFrom)
    GROUP BY participant.MeetingID
),0) as Value
UNION ALL
SELECT 'Tentative' as Response,
COALESCE((
    SELECT COUNT(*)
    FROM participant
    INNER JOIN invitation ON participant.InvitationID = invitation.InvitationID
    INNER JOIN invitationresponse ON invitation.InvitationResponseID =
invitationresponse.InvitationResponseID
    INNER JOIN responsetype ON invitationresponse.ResponseTypeID =
responsetype.ResponseTypeID
    WHERE participant.MeetingID = @ParamMeetingID
    AND participant.IsOrganizer = 0
    AND responsetype.Name = 'tentative'
    AND NOT ISNULL(invitation.DelegatedFrom)
    GROUP BY participant.MeetingID
),0) as Value
```

## 5.44 Delegations

Number of delegations per meeting. This is a similar metric to the number of re-plan request per meeting, but it is more passive. Discussion of the re-plan requests metric can be applied here.

**Number of delegeations in a specified meeting**

*@ParamMeetingID = ID of the specified meeting*

```sql
SELECT COUNT(*) as NumberOfDelegations
FROM participant
INNER JOIN invitation ON invitation.InvitationID = participant.InvitationID
WHERE participant.MeetingID = @ParamMeetingID
        AND NOT ISNULL(invitation.DelegatedTo)
GROUP BY participant.MeetingID
```

## 5.45 Delegations per person versus other employees

Delegation is one managerial habit. One metric for measuring management abilities of employees could be number of delegations versus the average delegation of other employees.

**Average delegation number in a specified interval (delegation number / invitation number)**

*@ParamStartTimeIntervalStart = start time of the specified „start time" interval*
*@ParamStartTimeIntervalEnd = end time of the specified „start time" interval*

```
SELECT COUNT(invitation.DelegatedTo)/COUNT(participant.InvitationID) as
NumberOfDelegatedMeetings
FROM participant
INNER JOIN meeting ON meeting.MeetingID = participant.MeetingID
INNER JOIN invitation ON invitation.InvitationID = participant.InvitationID
WHERE meeting.StartTime >= @ParamStartTimeIntervalStart
        AND meeting.StartTime <= @ParamStartTimeIntervalEnd
GROUP BY NULL
```

**Number of delegations per person versus average delegations from every employees in specified interval**

*@ParamEmployeeID = ID of the specified emplyoee*
*@ParamStartTimeIntervalStart = start time of the specified „start time" interval*
*@ParamStartTimeIntervalEnd = end time of the specified „start time" interval*

```
SELECT 'DelegationsFromThePerson' as Title,
COALESCE((
   SELECT COUNT(*) as NumberOfDelegatedMeetings
   FROM participant
   INNER JOIN meeting ON meeting.MeetingID = participant.MeetingID
   INNER JOIN invitation ON invitation.InvitationID = participant.InvitationID
   WHERE participant.EmployeeID = @ParamEmployeeID
         AND meeting.StartTime >= @ParamStartTimeIntervalStart
         AND meeting.StartTime <= @ParamStartTimeIntervalEnd
         AND NOT ISNULL(invitation.DelegatedTo)
   GROUP BY participant. EmployeeID
),0) as Value
UNION ALL
SELECT 'AverageDelegationsFromEveryEmployee' as Title,
COALESCE((
   SELECT COUNT(invitation.DelegatedTo)/COUNT(participant.InvitationID) as
NumberOfDelegatedMeetings
   FROM participant
   INNER JOIN meeting ON meeting.MeetingID = participant.MeetingID
   INNER JOIN invitation ON invitation.InvitationID = participant.InvitationID
   WHERE meeting.StartTime >= @ParamStartTimeIntervalStart
         AND meeting.StartTime <= @ParamStartTimeIntervalEnd
   GROUP BY NULL
),0) as Value
```

## 5.46 Decline after acceptance

This metric provides the number of declines per person, counting only those cases when a decline is sent after acceptance (introducing uncertainty by cancelling the previous acceptation).

**Number of „decline after acceptance" response from a specified person in meetings which are in a specified interval**

*@ParamEmployeeID = ID of the specified emplyoee*
*@ParamStartTimeIntervalStart = start time of the specified „start time" interval*
*@ParamStartTimeIntervalEnd = end time of the specified „start time" interval*

```
SELECT COUNT(*) as Result
FROM participant
INNER JOIN meeting ON meeting.MeetingID = participant.MeetingID
INNER JOIN invitation ON participant.InvitationID = invitation.InvitationID
INNER JOIN invitationresponse as current ON invitation.InvitationResponseID = current.InvitationResponseID
INNER JOIN invitationresponse as previous ON current.PreviousResponseID = previous.InvitationResponseID
INNER JOIN responsetype as currenttype ON current.ResponseTypeID = currenttype.ResponseTypeID
INNER JOIN responsetype as previoustype ON previous.ResponseTypeID = previoustype.ResponseTypeID
WHERE participant.EmployeeID = @ParamEmployeeID
    AND meeting.StartTime >= @ParamStartTimeIntervalStart
    AND meeting.StartTime <= @ParamStartTimeIntervalEnd
    AND currenttype.Name = 'decline'
    AND previoustype.Name = 'accept'
GROUP BY NULL
```

## 5.47 Last minute decline after acceptance

This metric provides the number of declines per person, counting only those cases when a decline is sent after acceptance, just before the meeting, this way introducing high uncertainty by cancelling the previous acceptation just before the meeting. Last minute range could be defined in minutes, depending on the organizational culture the threshold could be between 60 and 0.

**Number of „last minute decline after acceptance" response from a specified person in meetings which are in a specified interval**

Last minute: if „meeting start time" - „decline response date" < @ParamLastMinuteInterval

*@ParamEmployeeID = ID of the specified emplyoee*
*@ParamStartTimeIntervalStart = start time of the specified „start time" interval*
*@ParamStartTimeIntervalEnd = end time of the specified „start time" interval*
*@ParamLastMinuteInterval (in minute)*

```
SELECT COUNT(*) as Result
FROM participant
INNER JOIN meeting ON meeting.MeetingID = participant.MeetingID
INNER JOIN invitation ON participant.InvitationID = invitation.InvitationID
INNER JOIN invitationresponse as current ON invitation.InvitationResponseID = current.InvitationResponseID
INNER JOIN invitationresponse as previous ON current.PreviousResponseID = previous.InvitationResponseID
INNER JOIN responsetype as currenttype ON current.ResponseTypeID = currenttype.ResponseTypeID
INNER JOIN responsetype as previoustype ON previous.ResponseTypeID = previoustype.ResponseTypeID
WHERE participant.EmployeeID = @ParamEmployeeID
    AND meeting.StartTime >= @ParamStartTimeIntervalStart
    AND meeting.StartTime <= @ParamStartTimeIntervalEnd
```

```
    AND currenttype.Name = 'decline'
    AND previoustype.Name = 'accept'
    AND TIMESTAMPDIFF(MINUTE,current.DateTime,meeting.StartTime) <
@ParamLastMinuteInterval
GROUP BY NULL
```

## 5.48  Preparation time

It is often required that organizers and/or participants need time for preparation, for which they can allocate time in their calendars. If these preparation times could be linked to meetings, then additional costs of a meeting could also be better calculated. This metric can be calculated per person (e.g. average preparation time needed for meetings) or per meeting (sum of preparation time divided by number of participants). Training oriented meetings may require serious preparation times from the trainer and minimal preparation from attendees, co-working meetings may require similar preparation times from both the organizer and attendees.

**Total preparation time in a specified meeting**

*@ParamMeetingID = ID of the specified meeting*

```
SELECT SUM(PreparationTime)
FROM participant
WHERE MeetingID = @ParamMeetingID
GROUP BY NULL
```

**List of employees and their total preparation time in a specified interval**

*@ParamMeetingID = ID of the specified meeting*

```
SELECT participant.EmployeeID, SUM(PreparationTime)
FROM participant
INNER JOIN meeting ON meeting.MeetingID = participant.MeetingID
WHERE meeting.StartTime >= @ParamStartTimeIntervalStart
        AND meeting.StartTime <= @ParamStartTimeIntervalEnd
GROUP BY participant.EmployeeID
```

## 5.49  Meeting follow-up time

Similarly to the preparation time metric, follow-up time of a meeting could also be linked to the meeting and measured separately. Follow-up time includes all the time spent related to the meeting after the meeting. Follow-up time can include writing the minutes/summary of the meeting, including decisions of the meeting or next steps and tasks or it can include the rework and correction of errors (e.g. when a meeting was a review).

**Total follow-up time in a specified meeting**

*@ParamMeetingID = ID of the specified meeting*

```
SELECT SUM(FollowUpTime)
FROM participant
```

```
WHERE MeetingID = @ParamMeetingID
GROUP BY NULL
```
**List of employees and their total follow-up time in a specified interval**

*@ParamMeetingID = ID of the specified meeting*

```
SELECT participant.EmployeeID, SUM(FollowUpTime)
FROM participant
INNER JOIN meeting ON meeting.MeetingID = participant.MeetingID
WHERE meeting.StartTime >= @ParamStartTimeIntervalStart
        AND meeting.StartTime <= @ParamStartTimeIntervalEnd
GROUP BY participant.EmployeeID
```

## 5.50 Total time cost of a meeting

Total time cost of a meeting could be calculated as the sum of preparation time of participants, sum of meeting time of attendees and the sum of follow-up times. This could provide an overview how much effort a meeting requires. In calendar tools a suggestion could be added at planning to show an average effort needed for a meeting based on previous measurements at the company compared to the number of planned invitees. This could help in reducing the number of participants only to those who are really needed at the meeting.

**Total time cost in a specified meeting**

*@ParamMeetingID = ID of the specified meeting*

```
SELECT SUM(FollowUpTime + PreparationTime +
TIMESTAMPDIFF(HOUR,meeting.StartTime,meeting.FinishTime)) as TimeCost
FROM participant
INNER JOIN meeting ON meeting.MeetingID = participant.MeetingID
WHERE meeting.MeetingID = @ParamMeetingID
GROUP BY NULL
```
**List of employees and their total follow-up time in a specified interval**

*@ParamMeetingID = ID of the specified meeting*

```
SELECT participant.EmployeeID, SUM(FollowUpTime + PreparationTime +
TIMESTAMPDIFF(HOUR,meeting.StartTime,meeting.FinishTime)) as TimeCost
FROM participant
INNER JOIN meeting ON meeting.MeetingID = participant.MeetingID
WHERE meeting.StartTime >= @ParamStartTimeIntervalStart
        AND meeting.StartTime <= @ParamStartTimeIntervalEnd
GROUP BY participant.EmployeeID
```

## 5.51 Response and requiredness level

Distribution of responses for the invitation for "required" and "optional" groups. Acceptance shall be close to 100% for required invitations. The metric can be calculated per organization, per employee or per meeting.

**Number of acceptive/tentative/non-responding/decline number of invited participants (per requiredness level), who were invited to a specified meeting**

*@ParamMeetingID = ID of the specified meeting*

```
SELECT 'Acceptive - Required' as Response,
COALESCE((
   SELECT COUNT(*)
   FROM participant
   INNER JOIN invitation ON participant.InvitationID = invitation.InvitationID
   INNER JOIN invitationresponse ON invitation.InvitationResponseID = invitationresponse.InvitationResponseID
   INNER JOIN responsetype ON invitationresponse.ResponseTypeID = responsetype.ResponseTypeID
   WHERE participant.MeetingID = @ParamMeetingID
   AND participant.IsOrganizer = 0
   AND responsetype.Name = 'accept'
   AND invitation.Requiredness = 1
   GROUP BY participant.MeetingID
),0) as Value
UNION ALL
SELECT 'Decline - Required' as Response,
COALESCE((
   SELECT COUNT(*)
   FROM participant
   INNER JOIN invitation ON participant.InvitationID = invitation.InvitationID
   INNER JOIN invitationresponse ON invitation.InvitationResponseID = invitationresponse.InvitationResponseID
   INNER JOIN responsetype ON invitationresponse.ResponseTypeID = responsetype.ResponseTypeID
   WHERE participant.MeetingID = @ParamMeetingID
   AND participant.IsOrganizer = 0
   AND responsetype.Name = 'decline'
   AND invitation.Requiredness = 1
   GROUP BY participant.MeetingID
),0) as Value
UNION ALL
SELECT 'Non-Responding - Required' as Response,
COALESCE((
   SELECT COUNT(*)
   FROM participant
   INNER JOIN invitation ON participant.InvitationID = invitation.InvitationID
   WHERE participant.MeetingID = @ParamMeetingID
   AND participant.IsOrganizer = 0
   AND ISNULL(invitation.InvitationResponseID)
   AND invitation.Requiredness = 1
   GROUP BY participant.MeetingID
),0) as Value
UNION ALL
SELECT 'Tentative - Required' as Response,
COALESCE((
   SELECT COUNT(*)
   FROM participant
   INNER JOIN invitation ON participant.InvitationID = invitation.InvitationID
   INNER JOIN invitationresponse ON invitation.InvitationResponseID = invitationresponse.InvitationResponseID
   INNER JOIN responsetype ON invitationresponse.ResponseTypeID = responsetype.ResponseTypeID
   WHERE participant.MeetingID = @ParamMeetingID
   AND participant.IsOrganizer = 0
   AND responsetype.Name = 'tentative'
   AND invitation.Requiredness = 1
   GROUP BY participant.MeetingID
),0) as Value
UNION ALL
SELECT 'Acceptive - Optional' as Response,
COALESCE((
   SELECT COUNT(*)
   FROM participant
   INNER JOIN invitation ON participant.InvitationID = invitation.InvitationID
```

```
      INNER JOIN invitationresponse ON invitation.InvitationResponseID = 
invitationresponse.InvitationResponseID
      INNER JOIN responsetype ON invitationresponse.ResponseTypeID = 
responsetype.ResponseTypeID
      WHERE participant.MeetingID = @ParamMeetingID
      AND participant.IsOrganizer = 0
      AND responsetype.Name = 'accept'
      AND invitation.Requiredness = 0
      GROUP BY participant.MeetingID
),0) as Value
UNION ALL
SELECT 'Decline - Optional' as Response,
COALESCE((
      SELECT COUNT(*)
      FROM participant
      INNER JOIN invitation ON participant.InvitationID = invitation.InvitationID
      INNER JOIN invitationresponse ON invitation.InvitationResponseID = 
invitationresponse.InvitationResponseID
      INNER JOIN responsetype ON invitationresponse.ResponseTypeID = 
responsetype.ResponseTypeID
      WHERE participant.MeetingID = @ParamMeetingID
      AND participant.IsOrganizer = 0
      AND responsetype.Name = 'decline'
      AND invitation.Requiredness = 0
      GROUP BY participant.MeetingID
),0) as Value
UNION ALL
SELECT 'Non-Responding - Optional' as Response,
COALESCE((
      SELECT COUNT(*)
      FROM participant
      INNER JOIN invitation ON participant.InvitationID = invitation.InvitationID
      WHERE participant.MeetingID = @ParamMeetingID
      AND participant.IsOrganizer = 0
      AND ISNULL(invitation.InvitationResponseID)
      AND invitation.Requiredness = 0
      GROUP BY participant.MeetingID
),0) as Value
UNION ALL
SELECT 'Tentative - Optional' as Response,
COALESCE((
      SELECT COUNT(*)
      FROM participant
      INNER JOIN invitation ON participant.InvitationID = invitation.InvitationID
      INNER JOIN invitationresponse ON invitation.InvitationResponseID = 
invitationresponse.InvitationResponseID
      INNER JOIN responsetype ON invitationresponse.ResponseTypeID = 
responsetype.ResponseTypeID
      WHERE participant.MeetingID = @ParamMeetingID
      AND participant.IsOrganizer = 0
      AND responsetype.Name = 'tentative'
      AND invitation.Requiredness = 0
      GROUP BY participant.MeetingID
),0) as Value
```

## 5.52   Most optional employees

Top list of employees who are usually invited as optional. Top list can be calculated based on ratio of required/optional invitations. Employees with high optional invitations may not be decision makers.

**List of employees and their meeting numbers (where the employee is optional participant) in specified time interval. The list is ordered by the meeting numbers, first is the highest.**

*@ParamStartTimeIntervalStart = start time of the specified „start time" interval*
*@ParamStartTimeIntervalEnd = end time of the specified „start time" interval*

```
SELECT participant.EmployeeID, COUNT(*) as CountOfOptionalPerPerson
FROM participant
INNER JOIN meeting ON meeting.MeetingID = participant.MeetingID
INNER JOIN invitation ON participant.InvitationID = invitation.InvitationID
WHERE meeting.StartTime >= @ParamStartTimeIntervalStart
      AND meeting.StartTime <= @ParamStartTimeIntervalEnd
      AND invitation.Requiredness = 0
GROUP BY participant.EmployeeID
ORDER BY CountOfOptionalPerPerson DESC
```

### 5.53 Most required employees

Similarly to the most optional employees, this metric is the top list of employees who are usually invited as required. Top list can be calculated based on ratio of required/optional invitations. Employees with high required invitations and high number of meetings may be decision makers and may have influence within the organisation.

**List of employees and their meeting numbers (where the employee is required participant) in specified time interval. The list is ordered by the meeting numbers, first is the highest.**

*@ParamStartTimeIntervalStart = start time of the specified „start time" interval*
*@ParamStartTimeIntervalEnd = end time of the specified „start time" interval*

```
SELECT participant.EmployeeID, COUNT(*) as CountOfOptionalPerPerson
FROM participant
INNER JOIN meeting ON meeting.MeetingID = participant.MeetingID
INNER JOIN invitation ON participant.InvitationID = invitation.InvitationID
WHERE meeting.StartTime >= @ParamStartTimeIntervalStart
      AND meeting.StartTime <= @ParamStartTimeIntervalEnd
      AND invitation.Requiredness = 1
GROUP BY participant.EmployeeID
ORDER BY CountOfOptionalPerPerson DESC
```

### 5.54 Number of locations involved

Organization may have multiple sites, sometimes spread in different countries, with different areas of expertise and varying internal culture. Involving multiple locations into a meeting will face some barriers due to cultural, expertise, time zone etc. differences. The lower is the number of locations involved the easier can the meeting be organized.

**Number of locations in a specified meeting**

*@ParamMeetingID = ID of the specified meeting*

```
SELECT COUNT(LocationID)
FROM locationinmeeting
```

```
WHERE MeetingID = @ParamMeetingID
GROUP BY NULL
```
**List of meetings in a specified interval, which have more locations**

*@ParamStartTimeIntervalStart = start time of the specified „start time" interval*
*@ParamStartTimeIntervalEnd = end time of the specified „start time" interval*

```
SELECT meeting.MeetingID, COUNT(LocationID) as locations
FROM locationinmeeting
INNER JOIN meeting ON meeting.MeetingID = locationinmeeting.MeetingID
WHERE meeting.StartTime >= @ParamStartTimeIntervalStart
        AND meeting.StartTime <= @ParamStartTimeIntervalEnd
GROUP BY meeting.MeetingID
HAVING locations > 1
```

## 5.55  Delegations versus Delegation requests

Ratio of delegation requests received and delegation request sent out. Depending on the organization culture, this metric could show management (delegations sent) versus team player (delegations received) skills of an employee.

**Number of delegations versus delegation requests / person / specified interval**

*@ParamEmployeeID = ID of the specified emplyoee*
*@ParamStartTimeIntervalStart = start time of the specified „start time" interval*
*@ParamStartTimeIntervalEnd = end time of the specified „start time" interval*

```
SELECT 'DelegationsFromThePerson' as Title,
COALESCE((
   SELECT COUNT(*) as NumberOfDelegatedMeetings
   FROM participant
   INNER JOIN meeting ON meeting.MeetingID = participant.MeetingID
   INNER JOIN invitation ON invitation.InvitationID = participant.InvitationID
   WHERE participant.EmployeeID = @ParamEmployeeID
        AND meeting.StartTime >= @ParamStartTimeIntervalStart
        AND meeting.StartTime <= @ParamStartTimeIntervalEnd
        AND NOT ISNULL(invitation.DelegatedTo)
   GROUP BY participant. EmployeeID
),0) as Value
UNION ALL
SELECT 'DelegationRequestsForThePerson' as Title,
COALESCE((
   SELECT COUNT(*) as NumberOfDelegatedMeetings
   FROM participant
   INNER JOIN meeting ON meeting.MeetingID = participant.MeetingID
   INNER JOIN invitation ON invitation.InvitationID = participant.InvitationID
   WHERE participant.EmployeeID = @ParamEmployeeID
        AND meeting.StartTime >= @ParamStartTimeIntervalStart
        AND meeting.StartTime <= @ParamStartTimeIntervalEnd
        AND NOT ISNULL(invitation.DelegatedFrom)
   GROUP BY participant. EmployeeID
),0) as Value
```

# 6  Limitations

Metrics presented in this report are proposals which are needed to be validated in practice. Interpretation and usage of some of the metrics listed can depend on many parameters (e.g. company culture, situation or role) thus should be subjected to further research and should be used with precautions.

Metric presented in this report are raw results of a brainstorming, and were presented with the intention of raising the aswareness on the opportunity laying in calendar data. As such, the set of metrics were not checked for completeness, many more could be defined and used for various purposes.

Measurement of some of the metrics require sensitive data, e.g. those related to the employees and should always be used with the agreement of the employee.

# 7 Conclusion

In this report we proposed 50+ metrics for calendar mining with an argumentation on their possible usage. We also proposed a reference data model and queries in order to support better understanding of the metrics. Results presented in this report are intended to serve as a starting point for discussion possibilities of calendar mining. Hopefully the basic set of metrics will lead to the extension of the topic of calendar mining and can possibly inspire research and industry to use the data available behind calendar software. Options for further steps include definition of metrics for similar areas such as e-mail or chat mining[11], [12], refining the database and queries based on storing and querying real data, validating metrics by performing calendar mining at personal, team and organisational levels. In order to provide a solid basis for calendar mining, it would be useful to define a reference ontology[13] for harmonizing various calendar approaches. Another line of research could be the automatic analysis of meeting records (e.g. text mining[2], [3] and complexity analysis [14]). Industry can improve calendar software based on metrics proposed and support organisations with useful measurement data and enhanced their functionality (e.g. predictions, hints, restrictions and alerts); besides real-time tracking of meetings, such improvements in calendar software could help satisfying project planning and monitoring process requirements of improvement frameworks such as CMMI or SPICE [15]. Using concepts of process mining[1] could be used to understand processes related to meetings and to identify and eliminate related bottlenecks.

## Appendix A – Descriptions of tables in the data model

Table: meeting (contains meeting data)

**Attributes:**
- MeetingID: primary key (ID)
- InvitationTime: when the invitation has been sent
- StartTime: when the meeting is started

- FinishTime: when the meeting is finished
- RoomChangedNumber: how many times had the meeting room changed
- Cancelled: the meeting is cancelled or not
- RepetitionFrom:
    - If the meeting has one occurence (there are no repetitions), the RepetitionFrom attribute is equal to the MeetingID in this occurence
    - If the meeting has more occurences/repetitions, every occurence's RepetitionFrom attribute is equal to the first occurence's MeetingID
- NumberOfRescheduling: how many times was the meeting rescheduled

Table: location (contains location data)

**Attributes:**
- LocationID: primary key (ID)
- RoomID: foreign key to the room table (if the location is not a meeting room, this attribute is NULL)
- Label: if the location is not a meeting room, this free text contains the location name/description
- CompanyID: foreign key to the company table

Table: locationinmeeting (for storing the connection between location and meeting tables (N:M)).

**Attributes:**
- LocationInMeetingID: primary key (ID)
- MeetingID: foreign key to the meeting table
- LocationID: foreign key to the location table

Table: company (contains company data)

**Attributes:**
- CompanyID: primary key (ID)
- Name: Name of the company
- CountryID: foreign key to the country table

Table: country (contains country data).

**Attributes:**
- CountryID: primary key (ID)
- Name: name of the country

Table: employee (contains employee data)

**Attributes:**

- EmployeeID: primary key (ID)
- Name: employee's name
- Email: employee's e-mail address

Table: participant (contains meeting participant data)

**Attributes:**
- ParticipantID: primary key (ID)
- EmployeeID: foreign key to the employee table
- MeetingID: foreign key to the meeting table
- IsOrganizer: is this participants the organizer or not
- Invitation: foreign key to the invitation table (if this participant is non-invited, this attribute is NULL)
- ParticipateOnMeeting: this participant participated in the meeting or not
- PreparationTime: preparation effort
- FollowUpTime: follow up effort
- ModificationNumber: number of modifications from the participant

Table: invitation (contains invitation data)

**Attributes:**
- InvitationID: primary key (ID)
- Requiredness: the invitation acceptance is required or optional (value is true if required, false if optional)
- DelegatedTo: if the participant delegated the invitation to another participant, this attribute is the foreign key to the participant table (it can be NULL)
- DelegatedFrom: the participant who delegated the invitation to this participant (it can be NULL)
- InvitationResponseID: foreign key to invitationresponse table (to the last reponse)

Table: invitationresponse (contains invitation response data)

**Attributes:**
- InvitationResponseID: primary key (ID)
- DateTime: when the reponse has been sent
- ResponseTypeID: foreign key to the responsetype table
- PreviousResponseID: if this reponse is not the first in the current invitation, this attribute contains the ID of the previous response

Table: responsetype (contains response type data)

**Attributes:**

- ResponseTypeID: primary key (ID)
- Name: name of the response type

Table: inviteerequest (contains invitation response data)

**Attributes:**
- InviteeRequestID: primary key (ID)
- DateTime: when the request has been sent
- InvitationID: foreign key to the invitation table
- InviteeRequestTypeID: foreign key to the responsetype table

Table: inviteerequesttype (contains response type data)

**Attributes:**
- InviteeRequestTypeID: primary key (ID)
- Name: name of the request type

# Appendix B - SQL script for creating the database

```sql
SET @OLD_UNIQUE_CHECKS=@@UNIQUE_CHECKS, UNIQUE_CHECKS=0;
SET @OLD_FOREIGN_KEY_CHECKS=@@FOREIGN_KEY_CHECKS, FOREIGN_KEY_CHECKS=0;
SET @OLD_SQL_MODE=@@SQL_MODE, SQL_MODE='TRADITIONAL,ALLOW_INVALID_DATES';
-- -----------------------------------------------------
-- Schema mydb
-- -----------------------------------------------------
CREATE SCHEMA IF NOT EXISTS `mydb` DEFAULT CHARACTER SET utf8 ;
USE `mydb` ;

-- -----------------------------------------------------
-- Table `mydb`.`meeting`
-- -----------------------------------------------------
CREATE TABLE IF NOT EXISTS `mydb`.`meeting` (
  `MeetingID` INT NOT NULL AUTO_INCREMENT,
  `InvitationTime` DATETIME NOT NULL,
  `StartTime` DATETIME NOT NULL,
  `FinishTime` DATETIME NOT NULL,
  `RoomChangedNumber` INT NOT NULL DEFAULT 0,
  `Cancelled` TINYINT(1) NOT NULL DEFAULT 0,
  `RepetitionFrom` INT NULL,
  `NumberOfRescheduling` INT NOT NULL DEFAULT 0,
  PRIMARY KEY (`MeetingID`),
  INDEX `fk_meeting_meeting_idx` (`RepetitionFrom` ASC),
  CONSTRAINT `fk_meeting_meeting`
    FOREIGN KEY (`RepetitionFrom`)
    REFERENCES `mydb`.`meeting` (`MeetingID`)
    ON DELETE NO ACTION
    ON UPDATE NO ACTION)
ENGINE = InnoDB;

-- -----------------------------------------------------
-- Table `mydb`.`room`
-- -----------------------------------------------------
CREATE TABLE IF NOT EXISTS `mydb`.`room` (
  `RoomID` INT NOT NULL AUTO_INCREMENT,
  `Name` VARCHAR(45) NOT NULL,
  PRIMARY KEY (`RoomID`))
```

```sql
ENGINE = InnoDB;

-- -----------------------------------------------------
-- Table `mydb`.`country`
-- -----------------------------------------------------
CREATE TABLE IF NOT EXISTS `mydb`.`country` (
  `CountryID` INT NOT NULL AUTO_INCREMENT,
  `Name` VARCHAR(45) NOT NULL,
  PRIMARY KEY (`CountryID`))
ENGINE = InnoDB;

-- -----------------------------------------------------
-- Table `mydb`.`company`
-- -----------------------------------------------------
CREATE TABLE IF NOT EXISTS `mydb`.`company` (
  `CompanyID` INT NOT NULL AUTO_INCREMENT,
  `Name` VARCHAR(45) NOT NULL,
  `CountryID` INT NOT NULL,
  PRIMARY KEY (`CompanyID`),
  INDEX `fk_company_country1_idx` (`CountryID` ASC),
  CONSTRAINT `fk_company_country1`
    FOREIGN KEY (`CountryID`)
    REFERENCES `mydb`.`country` (`CountryID`)
    ON DELETE NO ACTION
    ON UPDATE NO ACTION)
ENGINE = InnoDB;

-- -----------------------------------------------------
-- Table `mydb`.`location`
-- -----------------------------------------------------
CREATE TABLE IF NOT EXISTS `mydb`.`location` (
  `LocationID` INT NOT NULL AUTO_INCREMENT,
  `Label` VARCHAR(45) NOT NULL DEFAULT 'No location information',
  `RoomID` INT NULL,
  `CompanyID` INT NOT NULL,
  PRIMARY KEY (`LocationID`),
  INDEX `fk_location_room1_idx` (`RoomID` ASC),
  INDEX `fk_location_company1_idx` (`CompanyID` ASC),
  CONSTRAINT `fk_location_room1`
    FOREIGN KEY (`RoomID`)
    REFERENCES `mydb`.`room` (`RoomID`)
    ON DELETE NO ACTION
    ON UPDATE NO ACTION,
  CONSTRAINT `fk_location_company1`
    FOREIGN KEY (`CompanyID`)
    REFERENCES `mydb`.`company` (`CompanyID`)
    ON DELETE NO ACTION
    ON UPDATE NO ACTION)
ENGINE = InnoDB;

-- -----------------------------------------------------
-- Table `mydb`.`employee`
-- -----------------------------------------------------
CREATE TABLE IF NOT EXISTS `mydb`.`employee` (
  `EmployeeID` INT NOT NULL AUTO_INCREMENT,
  `Name` VARCHAR(45) NOT NULL,
  `Email` VARCHAR(45) NOT NULL,
  PRIMARY KEY (`EmployeeID`))
ENGINE = InnoDB;

-- -----------------------------------------------------
-- Table `mydb`.`responsetype`
```

```sql
-- -----------------------------------------------------
CREATE TABLE IF NOT EXISTS `mydb`.`responsetype` (
  `ResponseTypeID` INT NOT NULL AUTO_INCREMENT,
  `Name` VARCHAR(45) NOT NULL,
  PRIMARY KEY (`ResponseTypeID`))
ENGINE = InnoDB;

-- -----------------------------------------------------
-- Table `mydb`.`invitationresponse`
-- -----------------------------------------------------
CREATE TABLE IF NOT EXISTS `mydb`.`invitationresponse` (
  `InvitationResponseID` INT NOT NULL AUTO_INCREMENT,
  `DateTime` DATETIME NOT NULL,
  `ResponseTypeID` INT NOT NULL,
  `PreviousResponseID` INT NULL,
  PRIMARY KEY (`InvitationResponseID`),
  INDEX `fk_invitationResponse_response1_idx` (`ResponseTypeID` ASC),
  INDEX `fk_invitationResponse_invitationResponse1_idx` (`PreviousResponseID` ASC),
  CONSTRAINT `fk_invitationResponse_response1`
    FOREIGN KEY (`ResponseTypeID`)
    REFERENCES `mydb`.`responsetype` (`ResponseTypeID`)
    ON DELETE NO ACTION
    ON UPDATE NO ACTION,
  CONSTRAINT `fk_invitationResponse_invitationResponse1`
    FOREIGN KEY (`PreviousResponseID`)
    REFERENCES `mydb`.`invitationresponse` (`InvitationResponseID`)
    ON DELETE NO ACTION
    ON UPDATE NO ACTION)
ENGINE = InnoDB;

-- -----------------------------------------------------
-- Table `mydb`.`invitation`
-- -----------------------------------------------------
CREATE TABLE IF NOT EXISTS `mydb`.`invitation` (
  `InvitationID` INT NOT NULL AUTO_INCREMENT,
  `Requiredness` TINYINT(1) NOT NULL,
  `DelegatedTo` INT NULL,
  `DelegatedFrom` INT NULL,
  `InvitationResponseID` INT NULL,
  PRIMARY KEY (`InvitationID`),
  INDEX `fk_invitation_participant1_idx` (`DelegatedTo` ASC),
  INDEX `fk_invitation_participant2_idx` (`DelegatedFrom` ASC),
  INDEX `fk_invitation_invitationresponse1_idx` (`InvitationResponseID` ASC),
  CONSTRAINT `fk_invitation_participant1`
    FOREIGN KEY (`DelegatedTo`)
    REFERENCES `mydb`.`participant` (`ParticipantID`)
    ON DELETE NO ACTION
    ON UPDATE NO ACTION,
  CONSTRAINT `fk_invitation_participant2`
    FOREIGN KEY (`DelegatedFrom`)
    REFERENCES `mydb`.`participant` (`ParticipantID`)
    ON DELETE NO ACTION
    ON UPDATE NO ACTION,
  CONSTRAINT `fk_invitation_invitationresponse1`
    FOREIGN KEY (`InvitationResponseID`)
    REFERENCES `mydb`.`invitationresponse` (`InvitationResponseID`)
    ON DELETE NO ACTION
    ON UPDATE NO ACTION)
ENGINE = InnoDB;

-- -----------------------------------------------------
-- Table `mydb`.`participant`
-- -----------------------------------------------------
CREATE TABLE IF NOT EXISTS `mydb`.`participant` (
```

```sql
  `ParticipantID` INT NOT NULL AUTO_INCREMENT,
  `EmployeeID` INT NOT NULL,
  `MeetingID` INT NOT NULL,
  `IsOrganizer` TINYINT(1) NOT NULL DEFAULT 0,
  `InvitationID` INT NULL,
  `ParticipateOnMeeting` TINYINT(1) NOT NULL DEFAULT 0,
  `PreparationTime` INT NOT NULL DEFAULT 0,
  `FollowUpTime` INT NOT NULL DEFAULT 0,
  `ModificationNumber` INT NOT NULL DEFAULT 0,
  PRIMARY KEY (`ParticipantID`),
  INDEX `fk_participant_employee1_idx` (`EmployeeID` ASC),
  INDEX `fk_participant_meeting1_idx` (`MeetingID` ASC),
  INDEX `fk_participant_invitation1_idx` (`InvitationID` ASC),
  CONSTRAINT `fk_participant_employee1`
    FOREIGN KEY (`EmployeeID`)
    REFERENCES `mydb`.`employee` (`EmployeeID`)
    ON DELETE NO ACTION
    ON UPDATE NO ACTION,
  CONSTRAINT `fk_participant_meeting1`
    FOREIGN KEY (`MeetingID`)
    REFERENCES `mydb`.`meeting` (`MeetingID`)
    ON DELETE NO ACTION
    ON UPDATE NO ACTION,
  CONSTRAINT `fk_participant_invitation1`
    FOREIGN KEY (`InvitationID`)
    REFERENCES `mydb`.`invitation` (`InvitationID`)
    ON DELETE NO ACTION
    ON UPDATE NO ACTION)
ENGINE = InnoDB;

-- -----------------------------------------------------
-- Table `mydb`.`inviteerequesttype`
-- -----------------------------------------------------
CREATE TABLE IF NOT EXISTS `mydb`.`inviteerequesttype` (
  `InviteeRequestTypeID` INT NOT NULL AUTO_INCREMENT,
  `Name` VARCHAR(35) NOT NULL,
  PRIMARY KEY (`InviteeRequestTypeID`))
ENGINE = InnoDB;

-- -----------------------------------------------------
-- Table `mydb`.`inviteerequest`
-- -----------------------------------------------------
CREATE TABLE IF NOT EXISTS `mydb`.`inviteerequest` (
  `InviteeRequestID` INT NOT NULL AUTO_INCREMENT,
  `DateTime` DATETIME NOT NULL,
  `InvitationID` INT NOT NULL,
  `InviteeRequestTypeID` INT NOT NULL,
  PRIMARY KEY (`InviteeRequestID`),
  INDEX `fk_inviteerequest_invitation1_idx` (`InvitationID` ASC),
  INDEX `fk_inviteerequest_inviteerequesttype1_idx` (`InviteeRequestTypeID` ASC),
  CONSTRAINT `fk_inviteerequest_invitation1`
    FOREIGN KEY (`InvitationID`)
    REFERENCES `mydb`.`invitation` (`InvitationID`)
    ON DELETE NO ACTION
    ON UPDATE NO ACTION,
  CONSTRAINT `fk_inviteerequest_inviteerequesttype1`
    FOREIGN KEY (`InviteeRequestTypeID`)
    REFERENCES `mydb`.`inviteerequesttype` (`InviteeRequestTypeID`)
    ON DELETE NO ACTION
    ON UPDATE NO ACTION)
ENGINE = InnoDB;

-- -----------------------------------------------------
-- Table `mydb`.`locationinmeeting`
```

```sql
-- -----------------------------------------------------
CREATE TABLE IF NOT EXISTS `mydb`.`locationinmeeting` (
  `LocationInMeetingID` INT NOT NULL AUTO_INCREMENT,
  `MeetingID` INT NOT NULL,
  `LocationID` INT NOT NULL,
  PRIMARY KEY (`LocationInMeetingID`),
  INDEX `fk_locationinmeeting_meeting1_idx` (`MeetingID` ASC),
  INDEX `fk_locationinmeeting_location1_idx` (`LocationID` ASC),
  CONSTRAINT `fk_locationinmeeting_meeting1`
    FOREIGN KEY (`MeetingID`)
    REFERENCES `mydb`.`meeting` (`MeetingID`)
    ON DELETE NO ACTION
    ON UPDATE NO ACTION,
  CONSTRAINT `fk_locationinmeeting_location1`
    FOREIGN KEY (`LocationID`)
    REFERENCES `mydb`.`location` (`LocationID`)
    ON DELETE NO ACTION
    ON UPDATE NO ACTION)
ENGINE = InnoDB;

SET SQL_MODE=@OLD_SQL_MODE;
SET FOREIGN_KEY_CHECKS=@OLD_FOREIGN_KEY_CHECKS;
SET UNIQUE_CHECKS=@OLD_UNIQUE_CHECKS;
```

## Appendix C HELPER: Parameters setting before queries

```sql
set @ParamMeetingID := 2;
set @ParamMeetingID1 := 1;
set @ParamMeetingID2 := 2;
set @ParamEmployeeID := 1;
set @ParamEmployeeID1 := 1;
set @ParamEmployeeID2 := 2;
set @ParamStartTimeIntervalStart := '2015.03.10';
set @ParamStartTimeIntervalEnd := '2015.09.26';
set @ParamCoreServiceStart := '9:00';
set @ParamCoreServiceFinish := '15:00';
set @ParamDateTime := '2015-11-21 13:00:00';
set @ParamRoomID = 1;
```

# Author CVs

### Zádor Dániel Kelemen

Zádor Dániel Kelemen received his BSc degree from "Gheorghe Asachi" Technical University of Iasi, Romania, the MSc degree from Budapest University of Technology and Economics, Hungary and his PhD degree from Eindhoven University of Technology, Netherlands, in the field of Software Process Improvement. He worked in various consultancy and leadership roles at SQI, ThyssenKrupp Presta and NNG. He attended thousands of meetings in his professional carreer which lead him to mine his own calendar and to define calendar mining metrics.

### Dániel Miglász

Dániel Miglász received his BSc degree from Budapest University of Technology and Economics, Hungary, he is currently an MSc student at Széchenyi István University, Hungary. He worked in various software developer roles at KBC Global Services Hungary, ThyssenKrupp Presta, Evosoft and Robert Bosch.